\documentclass[12pt]{article}

\usepackage[subpreambles=false]{standalone}

\usepackage[a4paper, margin=2cm]{geometry}
\usepackage[version=3]{mhchem} % Formula subscripts using \ce{}
\usepackage[numbers]{natbib} 
\usepackage[euler]{textgreek}
\usepackage{booktabs}
\usepackage{hyperref}
\usepackage{authblk}
\usepackage{rotating} % to rotate table
\usepackage{graphicx} % to load pdf as image
\usepackage{caption}
\usepackage{placeins}

\def\MIT{Massachusetts Institute of Technology}
\def\ChemE{Department of Chemical Engineering}
\def\CSE{Center for Computational Science and Engineering}
\def\MTL{Microsystems Technology Laboratories}
\def\MITaddress{77 Massachusetts Ave, Cambridge, MA, 02139, USA}
\def\DMSE{Department of Materials Science and Engineering}
\def\Nuclear{Department of Nuclear Science and Engineering}
\def\LL{Massachusetts Institute of Technology Lincoln Laboratory}
\def\LLaddress{244 Wood St, Lexington, MA, 02421, USA}

\def\sto{\ce{SrTiO3(001)}}
\def\gan{\ce{GaN(0001)}}
\def\si{\ce{Si(111)}}

\def\muga{\mu_{\text{Ga}}}
\def\musr{\mu_{\text{Sr}}}
\def\muti{\mu_{\text{Ti}}}
\def\muo{\mu_{\text{O}}}

\title{Machine-learning-accelerated simulations to enable automatic surface reconstruction}
\date{}
\author[1,2]{Xiaochen Du}
\author[2,3]{James K. Damewood}
\author[3]{Jaclyn R. Lunger}
\author[3]{Reisel Millan}
\author[3,4,5]{Bilge Yildiz}
\author[6]{Lin Li}
\author[3,*]{Rafael Gómez-Bombarelli}
\affil[1]{\ChemE, \MIT, \MITaddress}
\affil[2]{\CSE, \MIT, \MITaddress}
\affil[3]{\DMSE, \MIT, \MITaddress}
\affil[4]{\Nuclear, \MIT, \MITaddress}
\affil[5]{\MTL, \MIT, \MITaddress}
\affil[6]{\LL, \LLaddress}
\affil[*]{Corresponding author: Rafael Gómez-Bombarelli, rafagb@mit.edu}

\begin{document}

\maketitle

\begin{abstract}
    Understanding material surfaces and interfaces is vital in applications like catalysis or electronics. By combining energies from electronic structure with statistical mechanics, \textit{ab initio} simulations can in principle predict the structure of material surfaces as a function of thermodynamic variables. However, accurate energy simulations are prohibitive when coupled to the vast phase space that must be statistically sampled. Here, we present a bi-faceted computational loop to predict surface phase diagrams of multi-component materials that accelerates both the energy scoring and statistical sampling methods. Fast, scalable, and data-efficient machine learning interatomic potentials are trained on high-throughput density-functional theory calculations through closed-loop active learning. Markov-chain Monte Carlo sampling in the semi-grand canonical ensemble is enabled by using virtual surface sites. The predicted surfaces for \gan{}, \si{}, and \sto{} are in agreement with past work and suggest that the proposed strategy can model complex material surfaces and discover previously unreported surface terminations. 
\end{abstract}

\pagebreak

% \keywords{neural network force field, active learning, Markov-chain Monte Carlo, semi-grand canonical ensemble, density-functional theory}

\section{Introduction}\label{intro}
Surface structure determines the properties and performance of materials in application areas such as heterogeneous catalysis \cite{shi_recent_2017, sumaria_atomic-scale_2023}, electrocatalysis \cite{fabbri_dynamic_2017, zhang_hydrogen-induced_2022, sha_understanding_2023}, or electrochemical energy storage \cite{jung_understanding_2014, han_coating_2017, xu_bulk_2021}. Material surfaces are not just pristine cuts of the bulk structure, and even for a single surface facet, equilibrium reconstruction can lead to vastly different terminations and patterns as a function of temperature, external chemical potentials, and applied electrical potential \cite{hirata_electronic_1994, castell_scanning_2002, erdman_structure_2002, heifets_electronic_2007, li_data-driven_2023}. Experimental methods for studying surfaces at the atomic level are costly and involved so they cannot cover the wide range of experimental conditions \cite{merte_structure_2022, li_data-driven_2023}. 

Simulations have the potential to capture complex surface structures at a wide range of external conditions. To do so, accurate and computationally affordable surface energy predictions are needed, along with efficient statistical sampling across surface compositions and configurations. On the energy modeling front, while classical interatomic force fields with few fitted parameters are sufficient for simple surfaces such as those of Au and GaN \cite{foiles_embedded-atom-method_1986, nord_modelling_2003}, their simple functional forms are often unsuited for multi-component surfaces.

Accurate computational studies of complex surfaces have relied on energetics derived from expensive density-functional theory (DFT) simulations of human-input guess surfaces, but this strategy does not easily scale to the diversity of possible structures and cannot gather enough statistics for converging thermodynamic averages. Kolpak \textit{et al.} manually constructed candidate surfaces of \ce{BaTiO3}(001) with various coverage levels comprising vacancies and adsorbates of Ba, Ti, and O based on chemical intuition and previous experimental data \cite{kolpak_evolution_2008}. Then, they used DFT relaxations to produce an energy-based phase diagram that connected external conditions to surface structures. Nevertheless, human intuition does not guarantee sufficient exploration of the phase space to uncover the most thermodynamically stable structures, and cannot capture the role of entropy through single static structures \cite{wexler_automatic_2019, zhou_unexpected_2014, timmermann_iro2_2020}. 

Global optimization techniques such as basin hopping \cite{wales_global_1997, panosetti_global_2015, obersteiner_structure_2017, wexler_automatic_2019, egger_charge_2020, bauer_systematic_2022}, evolutionary algorithms \cite{wang_new_2014, bauer_systematic_2022}, random structure search \cite{schusteritsch_predicting_2014}, reinforcement learning \cite{meldgaard_structure_2020}, and simulated annealing \cite{hess_polar_2020} offer a principled approach to enhance the generalizability of such computational studies, enabling the discovery of unexpected phases in materials science. Nonetheless, these methods often entail trade-offs among the explored phase space, energy accuracy, and computational cost. In addition, the focus is on finding energy minima, not free energy minima. Schusteritsch and Pickard achieved a balance between energy accuracy and computational cost, but their exploration of compositions while investigating \ce{SrTiO3} grain boundaries was somewhat limited \cite{schusteritsch_predicting_2014}. Both Wexler \textit{et al.} and Wang \textit{et al.} managed extensive sampling across different compositions \cite{wang_new_2014, wexler_automatic_2019}. However, their reliance on computationally expensive DFT increased computational costs despite maintaining high energy accuracy. Meanwhile, Hess and Yildiz, in their study of \ce{La_{0.75}Sr_{0.25}MnO3}(001), reduced costs by utilizing electrostatic energy calculations derived from Bader charges, and limiting phase space using canonical Monte Carlo (MC) \cite{hess_polar_2020}. However, this lower-fidelity approach does not assure the stability of final structures.

Machine learning (ML) force fields are much faster than DFT calculations while preserving accuracy \cite{unke_machine_2021, axelrod_learning_2022} but it has been challenging to connect them to an efficient sampling scheme to explore the phase space of multi-component surfaces in an self-directed fashion. Recent works have introduced reconstruction pipelines leveraging ML force fields, although many require guess compositions and lack automatic consideration of external conditions in their sampling procedure \cite{bisbo_efficient_2020, bisbo_global_2022, merte_structure_2022, timmermann_iro2_2020, timmermann_data-efficient_2021, ronne_atomistic_2022, han_prediction_2023}. Furthermore, although ML-based sampling methods that vary surface compositions have demonstrated potential, they are currently limited to single adsorbate types on mono-component surfaces and may employ less efficient sampling schemes \cite{xu_atomistic_2022}.

We present our Automatic Surface Reconstruction (AutoSurfRecon) framework, which achieves thorough statistical sampling of thermodynamic states and avoids relying on hand-picked trial surfaces while being computationally efficient. AutoSurfRecon utilizes ML force fields and active learning (AL) to enable fast and accurate energetics. In addition, because surface reconstruction typically takes place on sites that emerge from the underlying slab, but not necessarily following slab symmetry, we utilize the computationally efficient Virtual Surface Site Relaxation-Monte Carlo (VSSR-MC) sampling algorithm in the semi-grand canonical ensemble, instead of the more intensive grand canonical MC. By populating virtual sites followed by relaxation, VSSR-MC samples across compositional and configurational search spaces to efficiently and accurately explore complex, stable surface structures across a range of external chemical potentials. To validate our sampling strategy based on virtual sites, we recover known reconstructions of the well-studied \gan{} and \si{} surfaces using classical force fields. We then show the effectiveness of the full AutoSurfRecon pipeline on \sto{}, a complex perovskite unsuited for classical force fields, using a neural network force field (NFF) energy model. We demonstrate that our uncertainty-driven AL strategy working in tandem with VSSR-MC sampling acquires new DFT data points only at relevant regions of the surface phase space. In fewer than 5000 DFT single-point calculations, the NFF can be trained to obtain accurate energy predictions for many distinct chemical compositions. Finally, we construct an \sto{} surface phase diagram that compares well with literature results and reveals unexpected low-energy surface terminations. Our algorithm can be easily applied to other surfaces of interest and we anticipate it can be used to study multi-component materials under challenging environments such as in aqueous electrochemistry.

\section{Results}\label{results}
\subsection{Development of the end-to-end automatic framework}\label{results:mcmc}
Fig.~1 shows our AutoSurfRecon computational workflow. Starting solely with a clean-cut surface and algorithmically generated virtual adsorption sites as inputs (see Section~\ref{methods:slab-modeling}), a surface phase diagram is produced. In this workflow, VSSR-MC sampled structures have the dual purpose of improving the NFF energy model and obtaining surface reconstruction statistics, further improving the efficiency of our pipeline. 

The phase diagram is obtained through successive surface reconstruction runs that, beginning with the pristine surface as input, explore the surface phase space using VSSR-MC. Runs are conducted at different elemental chemical potentials. 

Every individual VSSR-MC sampling iteration starts with discretely choosing one elemental identity at a selected adsorption site. The absorption sites are created algorithmically from the pristine slab geometry. A key innovation in VSSR-MC is that empty sites are defined as virtual atoms thus avoiding the higher volume space of grand canonical sampling in Cartesian coordinates. As such, adding an atom or removing an existing atom become instances of changing elemental identity and close contacts corresponding to very high repulsive energies can be easily rejected based on hard-sphere cutoffs. 

Following discrete sampling, all adsorbates and surface atoms undergo continuous relaxation of atomic positions away from their assigned sites (see Section~\ref{methods:mcmc}), allowing us to efficiently explore the vast configurational space as lower-energy relaxations of easy-to-sample virtual sites. The energy model used is a classical force field for mono- and bi-component materials or a machine-learned force field model for complex materials. The structure obtained after discrete sampling and continuous relaxation is evaluated with the acceptance criterion corresponding to the semi-grand canonical ensemble \cite{bernardin_semi-grand_2007, damewood_sampling_2022}:
\begin{align}
    P &= \min{\left\{1, \exp\left(-\frac{\Delta E_{\text{slab}}-\Delta \mu}{k_B T}\right)\right\}} \label{equation:semigrand_accept_criteria}
\end{align}
where $P$ is the probability of acceptance, $\Delta E_{\text{slab}}$ is the change in slab energy after both discrete sampling and continuous relaxation, $\Delta \mu$ is the change in chemical potential due to discrete sampling, $k_B$ is the Boltzmann constant, and $T$ is the sampling temperature. 

We also developed an AL strategy (Fig.~2) to efficiently acquire the fewest possible DFT-evaluated structures required for NFF training. At each AL iteration, we train an ensemble of NFF models for uncertainty quantification \cite{carrete_deep_2023, tan_single-model_2023} (see Section \ref{methods:nnff}). To iteratively select structures for DFT evaluation, we use uncertainty-based adversarial attack \cite{schwalbe-koda_differentiable_2021} or VSSR-MC combined with latent space clustering to seek out structures that maximize NFF prediction error, which in practice is estimated by the predicted force standard deviation (SD) (see Section \ref{methods:al}). The adversarial attack algorithm displaces atomic positions of existing structures to regions of high SD but not high energy \cite{schwalbe-koda_differentiable_2021}, thereby improving the accuracy and stability of the force field \cite{fu_forces_2022}. However, adversarial attacks cannot autonomously sample different compositions. VSSR-MC samples both new compositions and configurations across chemical potentials and guides the NFF to learn only relevant subsets of the vast phase space. At the same time, clustering candidate structures reduces the number of training data by only selecting those whose local chemical environments are unique \cite{damewood_representations_2023}.

\subsection{Reconstructions with classical potentials}
To demonstrate our VSSR-MC sampling method, we investigated the 3x3 reconstruction of the \gan{} surface using a Tersoff potential \cite{nord_modelling_2003} and the 3x3, 5x5, and 7x7 reconstructions of the \si{} surface using a modified Stillinger-Weber potential \cite{stephenson_modified_1996}.

\subsubsection{\gan{} reconstruction}\label{results:classical:GaN}
\ce{GaN} is a well-studied semiconductor and the (0001) surface was described to have a contracted adsorption layer on a rotated lattice with respect to the pristine surface \cite{northrup_structure_2000}. When running the VSSR-MC algorithm at a fixed number of Ga adsorbates, several reconstructions matching the literature were obtained (Fig.~3). When viewed from the top, our structure in Fig.~3(a) shows the same rhombus patterns as the literature structure. The adsorbate distance from the pristine surface layer matched the literature value of 2.42 Å (side view, Fig.~3(b)) and the energy difference stood at a mere 0.008 meV/atom. The energy approached the ground state and VSSR-MC acceptance rate neared 0 at around 20,000 to 25,000 iterations in Fig.~3(c). Additional reconstructed surfaces can be found in Extended Data Fig.~1.

\subsubsection{\si{} reconstructions}\label{results:classical:Si}
\si{} is known to exhibit complex dimer-adatom stacking fault (DAS) surface reconstructions that vary with supercell size \cite{stich_ab_1992,smeu_electronic_2012}. These reconstructions neither follow bulk lattice geometries nor virtual site geometries. We separately investigated the 3x3, 5x5, and 7x7 surfaces and recovered the DAS reconstructions in our VSSR-MC runs at a fixed adsorbate number for each supercell (Extended Data Fig.~2). Multiple DAS-like structures that are within the thermally-accessible energy range (25.7 meV/atom) at room temperature were obtained for all supercell sizes, two of such 3x3 structures are shown in Extended Data Fig.~2(a) and one of each for 5x5 and 7x7 is presented in Extended Data Fig.~2(b) and (c) respectively. Moreover, VSSR-MC also sampled the 5x5 pristine surface, which is within 2.1 meV/atom of the DAS structure according to the potential, as well as various 7x7 pristine-like structures with four adatoms (due to stoichiometry differences) that are 3.5 meV/atom higher in energy than the DAS target. By discovering these unexpected low-energy structures alongside DAS structures, VSSR-MC demonstrated the ability to thoroughly sample a large phase space of surface reconstructions.

\subsection{\sto{} reconstructions with neural force field}\label{results:STO}
We validated the performance of our full AutoSurfRecon framework on a challenging surface for which there is no known analytical potential. We chose \sto{} because \ce{SrTiO3} is representative of the complex perovskite oxide family and the (001) surface, in particular, is stable and demonstrates a variety of surface reconstruction patterns under different elemental chemical potentials \cite{hirata_electronic_1994, erdman_structure_2002, castell_scanning_2002, heifets_electronic_2007}.

\subsubsection{Active learning for neural network force field}\label{results:STO:AL}
Including the initial dataset, AL was run for a total of six iterations (see Section~\ref{methods:al} for details) on \sto{} slabs of varying compositions (example in Fig.~4(a)). A total of 6500 structures were selected for DFT evaluation: the first and last AL generations resulted in about 1500 structures each while AL generations 2-5 yielded approximately 800 structures each. Across our AL runs, we found a good correlation between force mean absolute error (MAE) and predicted force SD, demonstrating the validity of our error estimation procedure. Fig.~4(b) shows one such correlation plot derived from the final NFF model and sixth-generation structures. Correlation plots across all generations are provided in Extended Data Fig.~3.

Fig.~4(c) shows a principal component analysis (PCA), 2D projection of all new surface structures generated through our AL process. In order to create a consistent representation across generations, PCA was applied to neural network embeddings from the last-generation model on the full dataset. In Fig.~4(c), we see a distinct pattern of VSSR-MC structures compared with the initial structures, with VSSR-MC structures evolving over generations 3-5 to show streaks. These patterns suggest with NFF improvement, VSSR-MC was able to more effectively sample structures corresponding to surface energy minima. We also observe a correlation of the predicted energy with the first two principal axes (see training objective in Section~\ref{methods:nnff}). Additional information on the distribution of forces obtained at each AL generation is provided in Extended Data Fig.~4.

Fig.~4(d) shows the performance of the NFF model, measured by energy MAE and force MAE, after each AL iteration over a common test data. The force and energy MAE drop substantially within four AL iterations, suggesting the effectiveness of the AL process for sampling the most informative surfaces for ML model training. While the overall trend of both MAEs is down, the increase in energy MAE during generation 2 can be attributed to the distribution mismatch between the starting structures and the MC-sampled structures, suggesting the first-generation model was overfit to the initial dataset and struggled to generalize to the much broader and realistic phase space explored during the AL cycles. The decrease from generation 3 to 4 was the most dramatic and in generation 4, an improved NFF allowed VSSR-MC samples to more closely follow the underlying distribution, resulting in the performance improvement. 

Overall, after data splitting, the final NFF was fitted on a training set with fewer than 5000 structures and achieved a force MAE of 0.10 eV/Å and an energy MAE of 5.18 meV/atom (see Extended Data Fig.~5) across the phase space needed to power production VSSR-MC runs. It is important to note that the number of AL generations is not fixed, but is determined by the observed plateau in NFF accuracy improvements and the convergence of VSSR-MC structures. The goal of our AL methodology is to optimize computational resources while attaining good prediction accuracy for a diverse set of structures. Future research may include the exploration of more precise or automated stopping criteria, such as the implementation of force or energy thresholds.

\subsubsection{Data analysis and constructing the phase diagram}
We first present a summary of literature data on \sto{} reconstructions. The structure of \sto{} has inspired intensive research and debate. A double-layer (DL) \ce{TiO2}-terminated surface is supported by both experiments and theory \cite{erdman_structure_2002, herger_surface_2007, hong_anomalous_2023}. But single-layer (SL) \ce{TiO2} and SL \ce{SrO} terminations have also been reported \cite{castell_scanning_2002, hirata_electronic_1994, szot_surfaces_1999, kubo_surface_2003}. All these surfaces can be related to the chemical potential of the constitutive elements in the material bulk, which is assumed to be in equilibrium with the environment. In this case, the Sr chemical potential, $\musr$, is the most relevant. By increasing $\musr$, which can be achieved by evaporating Sr metal into the reaction environment, adding Sr to the surface becomes more favorable, resulting in the depletion of Ti and finally the formation of an SrO adlayer \cite{hirata_electronic_1994, heifets_electronic_2007}. 

In our surface reconstruction runs, we observed all three terminations starting from DL \ce{TiO2} in low $\musr$ to SL \ce{TiO2} (that is, the unreconstructed surface) in intermediate $\musr$, and SL \ce{SrO} in high $\musr$. A schematic is provided in Fig.~5(a) to illustrate this relationship. Given our choice of modeling \sto{}, the pristine surface mirrored the stoichiometry of the bulk. As such, alterations to $\musr$ influenced surface free energies in the same manner as equivalent changes in $\muo$, the O chemical potential, eliminating the need for additional runs varying $\muo$ (as referenced in Equation~\ref{equation:final-surf}).

Analyzing our data obtained from VSSR-MC, we constructed a surface phase diagram of \sto{} in Fig.~5(b) that maps chemical potentials of Sr and O to the most stable surface terminations. Our phase diagram matches the expected trend in $\musr$. It is also similar to the one obtained by Heifets \textit{et al.} with a narrow strip of SL \ce{TiO2} phase sandwiched between a DL \ce{TiO2} phase at substantial oxygen vacancies (low $\muo$) and low $\musr$, and a SL \ce{SrO} phase at considerable Sr concentrations in an oxygen atmosphere (high $\muo$ and $\musr$) \cite{heifets_electronic_2007}. We additionally assign three experimental \sto{} surfaces to our phase diagram, taking into account $\musr$ is loosely related to experimental procedures while $\muo$ can be calculated from \textit{p}\textsubscript{\ce{O2}}, the partial pressure of \ce{O2} gas, and the experimental temperature. 

To construct the phase diagram, the surface free energy ($\Omega_{\text{surf}}$) was recalculated for each structure using the final NFF model (see Section~\ref{methods:surface_free_energy}) at various $\musr$ and we plotted $\Omega_{\text{surf}}$ against the difference in the number of Sr and Ti atoms ($\Gamma^{\text{Ti}}_{\text{Sr}}$) for each slab. In the plots for $\musr = -10$ eV in Fig.~5(c), $\musr = -7$ eV in Fig.~5(d), and $\musr = -4$ eV in Fig.~5(e), we see that structures near the minimum $\Omega_{\text{surf}}$ correspond to the three known terminations. DL \ce{TiO2} (Fig.~5(c)) and SL \ce{TiO2} (Fig.~5(d)) had the lowest energy at $\musr = -10$ eV and $\musr = -7$ eV respectively. SL \ce{SrO} (Fig.~5(e)) had the second lowest energy at $\musr = -4$ eV, and became the lowest after filtering for stoichiometry.

\subsubsection{Comparing double-layer \ce{TiO2} terminations}
We additionally show that VSSR-MC faithfully recreated different reconstruction patterns of the DL \ce{TiO2} surface. As reported by \cite{erdman_structure_2002, herger_surface_2007} and others, DL \ce{TiO2} does not consist of a single termination. 2x2, 2x1 and 1x1 terminations are possible and we describe these terminations in Fig.~6. The most common 2x2 termination in literature is denoted in Fig.~6 as 2x2-A. The dominant DL \ce{TiO2} terminations could vary based on the exact surface structure and exchange-correlation functional, calculation settings, and empirical dispersion or Hubbard corrections. In this case, VSSR-MC samples contain two out of the three literature terminations: 2x2-A and 1x1. These two \ce{TiO2} terminations are close to one another in stability, as in \cite{herger_surface_2007}. 

The algorithm also discovered two surface terminations not previously reported, which we denote as 2x2-B and 2x2-C. 2x2-C has a similar energy to the two observed literature terminations but 2x2-B is lower in energy. The 2x1 reconstruction reported in the literature was not observed during surface reconstruction runs. Energy prediction using NFF and confirmed using DFT demonstrate an energy more than 2 eV above that of the most stable surface (2x2-B) and almost 1 eV above that of the next highest energy termination (2x2-C). We also observe the consistency between NFF predictions and DFT energies, which again shows the accuracy of NFFs as an energy model.

\section{Discussion}\label{discuss}
The presented algorithm overcomes limitations of previous computational methods, as shown in Extended Data Table 1. We anticipate VSSR-MC will be broadly applicable and aim to extend it to more challenging multi-component solid-liquid interfaces under electrical potential. 

VSSR-MC is advantageous because its trials are limited to high-likelihood virtual sites responsible for reconstructions that relate to the symmetry of the pristine slab. A disadvantage is that it is focused on thin, periodic reconstructions and may struggle to reconstruct amorphized thicker slabs that do not follow virtual site geometries. A schematic on the strengths and limitations of our sampling approach is given in Extended Data Fig.~6.

In this work, we sampled only one adsorption layer of a complex oxide. Future work would benefit from an improved understanding of complex surface reconstructions as results of stoichiometric changes across multiple adsorption layers, or to vacancies in the bulk. 

While our approach is not strictly constrained by the size of the unit cell, for \sto{} especially, we chose to use a relatively small unit cell whereas larger ones could show more complex reconstruction patterns. Moving to a larger unit cell could require transfer learning from NFFs trained on existing data. Well-trained NFFs typically generalize well to larger symmetry-breaking supercells \cite{winter_simulations_2023, millan_effect_2023}. As with other methods for studying surface reconstruction, the choice of the unit cell should be informed by past experimental and computational studies. 

As is common in the field, we approximated the surface free energy directly from DFT energies. However, there could be instances where vibrational contributions to the surface free energy can be important, especially at higher temperatures. The speed of NFFs would in principle allow adding free-energy corrections based on the harmonic approximation at tractable cost. 

Additionally, there is a wide literature of \sto{} reconstructions and not all past studies agree with each other. Further experimental studies probing different combinations of chemical potentials will help validate our phase diagram. 

Finally, we acknowledge that Markov chain-based sampling is difficult to parallelize and we envision an ML-based sampling method in the future to improve sampling speed.

\section{Methods}\label{methods}

\subsection{Virtual Surface Site Relaxation-Monte Carlo}\label{methods:mcmc}
The Metropolis-Hastings Markov-chain Monte Carlo algorithm was adapted to implement VSSR-MC. VSSR-MC simulations were performed in the canonical ensemble in addition to the semi-grand ensemble. In the canonical ensemble simulation, the number of each adsorbate type is fixed. For each canonical VSSR-MC iteration, a pair of adsorption sites with different adsorbate types (empty virtual sites count as one adsorbate type) are randomly chosen and the adsorbate identities are swapped. In the semi-grand ensemble simulation, the chemical potentials of adsorbates are supplied and the total number of adsorbates may vary across a MC run. For each semi-grand VSSR-MC iteration, one adsorption site is randomly chosen to change state.

For \gan, the semi-grand ensemble method was first used with $\muga=5$ eV (arbitrary positive value) to increase the number of Ga adsorbates to 12 before switching over to the canonical ensemble for annealing with annealing parameter $\alpha=0.99$. The starting sampling temperature was varied between 5000 K and 12,000 K (temperature at which $k_B T$ = 1).

For \si, a similar procedure was followed for canonical ensemble sampling with $\alpha=0.99$ and starting temperature at 12,000 K. Low-temperature constant-temperature sampling varying between 300 K and 2500 K was also employed and was more effective at converging the simulations for the 5x5 and 7x7 supercell reconstructions. The 3x3, 5x5, and 7x7 supercells had 16, 50, and 102 Si adsorbates respectively. For the larger \si{} 5x5 and 7x7 supercells, swaps between closer virtual sites were favored using exponentially-decaying distance weights:
\begin{align}
    w_{ij} &= \frac{\exp{(-d_{ij}/d_0)}}{\displaystyle\sum_{\text{all } j} \exp{(-d_{ij}/d_0)}}
\end{align}
where $w_{ij}$ is the normalized weight between the first selected virtual site $i$ and the second selected virtual site $j$, $d_{ij}$ is the Euclidean distance between $i$ and $j$, and $d_0$ is a user-defined distance decay factor that was set to 2.35 Å, the nearest-neighbor distance in Si bulk. For the 5x5 and 7x7 surfaces, multiple DAS-like structures within the thermal energy threshold were sampled in a single MC chain. Including these low-energy structures as seeds in subsequent VSSR-MC runs produced the DAS structures. 

For \sto, semi-grand VSSR-MC was run at various $\musr$ and a sampling temperature of at least 1000 K. (See \ref{methods:al} for details.) The chemical potential of empty virtual surface sites is set to 0 eV.

Following the discrete sampling step, continuous relaxation was performed using the conjugate gradient method in LAMMPS \cite{thompson_lammps_2022} (for the adsorbate atoms in \gan{} and both surface and adsorbate atoms in \si{}) or the BFGS algorithm in ASE \cite{larsen_atomic_2017} (for both surface and adsorbate atoms in \sto). For the \gan{} and \si{} 3x3 surfaces, a maximum of 500 relaxation steps was allowed after each discrete sampling step. For the \si{} 5x5 and 7x7 surfaces, a maximum of 100 relaxation steps was set to save on computational cost. For \sto{}, a maximum ranging from 5 to 20 steps in 5-step increments was allowed to reduce computational cost and to capture non-zero forces for AL.

For each iteration in the canonical ensemble, the acceptance probability $P$ is given by the minimum of unity and the ratio of the Boltzmann weights between the proposed and current state:
\begin{align}
    P &= \min{\left\{1, \exp\left(-\frac{\Delta E_{\text{slab}}}{k_B T}\right)\right\}}
\end{align}
where $\Delta E_{\text{slab}}$ is the change in slab energy after both discrete and continuous sampling, $k_B$ is the Boltzmann constant, and $T$ is the sampling temperature. 

\subsection{Surface slab modeling}\label{methods:slab-modeling}
The Python ASE, CatKit \cite{boes_graph_2019}, and pymatgen \cite{ong_python_2013} libraries were used to create and manipulate surfaces as well as to generate virtual adsorption sites defined from the bulk. Both pymatgen \verb|pymatgen.analysis.adsorption.AdsorbateSiteFinder.find_adsorption_sites| and CatKit \verb|catkit.gen.adsorption.get_adsorption_sites| produce well-covered, visually-dense top, bridge, and hollow adsorption sites with minimal changes from default settings. Within CatKit, adsorption sites are defined on the same plane as the surface and depend on another method to adsorb atoms at the correct distance from the surface. Converting CatKit coordinates to virtual sites for use in VSSR-MC required additional steps. Thus, for \si{} and \sto{}, we only used pymatgen virtual sites. The \verb|AdsorbateSiteFinder| class creates adsorption sites using a Delaunay triangulation-based algorithm. The analogous CatKit method also uses a geometry-based method to generate adsorption sites. VESTA was used for visualization and producing figures \cite{momma_vesta_2011}.

\subsubsection{\gan{} slab modeling}
A \ce{GaN} hexagonal unit cell from the Materials Project \cite{jain_commentary_2013} (mp-804) was cut in the (0001) plane to form a 3x3 supercell with two layers. A vacuum spacing of 15 Å in total was added to the ends of the slab. Both pymatgen- and CatKit-generated virtual sites worked equally well. Setting the symmetry reduce option to False to generate more virtual sites was the only change made to the default settings for both methods. A side view of the pymatgen and CatKit virtual sites can be found in Extended Data Fig.~7(a-b). For this surface, Ga was the only adsorbate.

\subsubsection{\si{} slab modeling}
A \ce{Si} cubic unit cell from the Materials Project (mp-149) was cut in the (111) plane to form 3x3, 5x5, and 7x7 supercells with four layers each comprising 9, 25, and 49 atoms respectively. A vacuum spacing of 20 Å in total was added to the ends of the slab. pymatgen sites with up to two layers were employed in separate MC runs. For a single layer, the adsorption sites were defined at 3.0 Å, the expected distance of adsorbates from the surface. For two layers, the bottom layer ranged from 2.0 to 3.0 Å in 0.5 Å increments and the top layer ranged from 3.5 to 5.0 Å also in 0.5 Å increments. A single layer was sufficient to sample structures within the thermally-accessible window and the pristine(-like) 5x5 and 7x7 surfaces. Similar to \gan{}, symmetry reduction of sites was disabled. A side view of the 5x5 supercell pymatgen virtual sites can be found in Extended Data Fig.~7(c). For this surface, Si was the only adsorbate.

\subsubsection{\sto{} slab modeling}
An \ce{SrTiO3} cubic unit cell from the Materials Project (mp-5229) was optimized and cut in the (001) plane to create a 2x2 supercell with three layers of \ce{TiO2} and SrO. A vacuum spacing of 15 Å in total was set at the ends of the slab. Overlapping adsorption sites (${\sim}{100}$) were defined at an arbitrary distance of 1.55 Å from the \ce{TiO2} surface (cf. 1.96 Å lattice parameter in the unit cell) using the same pymatgen \verb|AdsorbateSiteFinder| class. Testing different adsorption site distances up to 1.7 Å in early runs showed adsorbates consistently relaxed to around the same distance away from the pristine surface. Symmetry reduction of sites was similarly disabled. A side view of the pymatgen virtual sites can be found in Extended Data Fig.~7(d). For this surface, Sr, Ti, and O were possible adsorbates.

\subsection{Neural network force field}\label{methods:nnff}
NFFs for \sto{} were trained using an internal implementation of the equivariant PaiNN architecture \cite{schutt_equivariant_2021}. The NFF energies are not the same as surface free energies (as in Section~\ref{methods:surface_free_energy}). A linear interpolation was performed over the lowest DFT energy structure for each composition to derive the energy offset per atom type. The corresponding atomic energy offset was subtracted from each atom in all structures to produce the target NFF energies. The original network hyperparameters were determined with SigOpt \cite{martinez-cantin_practical_2018} to provide good results and thus unmodified. Briefly, the interaction cutoff between atoms was 5.0 Å, feature dimension was 128, number of radial basis functions was 20, and number of convolutions, 3. The swish activation function was employed \cite{ramachandran_searching_2017}. The loss function was a weighted sum of the mean-squared errors of forces and energy in the 100:1 ratio. The Adam optimizer \cite{kingma_adam_2015} was used with a starting learning rate (LR) of 0.001, an LR patience of 15 steps, and an LR decay factor of 0.3. A stopping potential in the form of $V = (\frac{\sigma}{r})^{12}$ was also employed. An ensemble of three NFF models was trained for 500 epochs each, sufficient to yield reliable estimates of energy, forces, and their respective variances.

\subsection{Active learning}\label{methods:al}
To create the starting \ce{SrTiO3}(001) dataset, Sr, Ti, and O adsorbates were randomly added to adsorption sites, using a minimum distance of 1.5 Å between atoms to prevent non-physical results.

The adversarial attack method, as implemented by Schwalbe-Koda \textit{et al.} \cite{schwalbe-koda_differentiable_2021}, centers around the adversarial loss function, $\mathcal{L}_{\text{adv}}$, which maximizes the predicted Boltzmann probabilities, $p(X_\delta)$, and the variance of predicted forces, $\sigma_F^2(X_\delta)$, via direct nuclear coordinate perturbation:

\begin{align}
\underset{\delta}{\max}\:\mathcal{L}_{\text{adv}}(X, \delta; \theta) &= \underset{\delta}{\max}\:p(X_\delta)\sigma_F^2(X_\delta)
\end{align}

In this equation, $X$ denotes initial atomic positions, $\delta$ represents atomic displacements from these positions, and $\theta$ signals the use of a neural network to estimate $\mathcal{L}_{\text{adv}}$. The predicted Boltzmann probabilities, $p(X_\delta)$, are proportional to $\exp{\left(\frac{-E_{\text{slab}}(X_\delta)}{k_BT_{\text{adv}}}\right)}$, where $E_{\text{slab}}(X_\delta)$ is the energy of the displaced slab, $k_B$ is the Boltzmann constant, and $T_{\text{adv}}$ is the adversarial attack sampling temperature. Both $p(X_\delta)$ and $\sigma_F^2(X_\delta)$ are estimated by our NFF. The atomic coordinates of each atom are perturbed independently to maximize the adversarial loss function. The adversarial attack was performed using default parameters, with $k_BT_{\text{adv}}$ set at 0.7 eV, LR at $5\times 10^{-4}$, and run for 100 epochs. For the second-generation structures, starting structures (seeds) were selected randomly, while for the sixth-generation structures, seeds with energies up to 43.4 eV (1000 kcal/mol) above the lowest energy structure were chosen.

AL using latent space clustering was done by first running VSSR-MC sampling from $\musr=-12$ eV to $\musr=0$ eV in increments of 2 eV. The MC-generated structures were clustered according to the first three principal components (${\sim}90\%$ explained variance) of their NFF embeddings and the most uncertain structure for each cluster was selected. The predicted force standard deviation, rather than energy SD, was used to estimate uncertainty, as it more accurately estimates the NFF error as seen in this work and previous research \cite{gasteiger_fast_2020, schwalbe-koda_differentiable_2021}. Our approach applies hierarchical agglomerative clustering with Ward's method. This technique initially treats each data point as a separate cluster and then progressively merges clusters, minimizing total within-cluster variance. To save on compute time, a first-pass clustering was run for every 1000 samples to yield around 100-200 structures at each $\musr$. The results of one such clustering run is shown in Extended Data Fig.~8(a). Structures sampled at all $\musr$ were pooled together for a second clustering step to select around 800 structures per AL generation.  

Including starting structures, AL was run for a total of six generations. The second AL generation used adversarial attack to create a stable NFF for continuous relaxation. The next three generations (3-5) employed VSSR-MC with latent space clustering to train our NFF in the phase space most relevant to surface reconstruction. The goal of using adversarial attack for the final AL generation was to make the force field more robust for low-energy structures.

\subsection{Surface stability analysis}\label{methods:surface_free_energy}
The analysis method is outlined below, details are in Supplementary Section 2. \sto{} surfaces were compared using the surface Gibbs free energy $\Omega^{\ce{SrTiO3}}_{\text{surf}}$ \cite{heifets_electronic_2007}:
\begin{align}
    \Omega^{\ce{SrTiO3}}_{\text{surf}} &= G_{\text{slab}} - N_{\text{Sr}}\musr - N_{\text{Ti}}\muti - N_{\text{O}}\muo
\end{align} where $G_{\text{slab}}$ refers to the Gibbs free energy of the slab. For each element $a$, $N_{a}$ refers to the number of $a$ atoms in the slab, $\mu_a$ refers to the chemical potential of $a$.

By the following relationship: 
\begin{align}
    \mu_{\text{\ce{SrTiO3}}} &= \musr + \muti + 3 \muo = g^{\text{bulk}}_{\text{\ce{SrTiO3}}} \nonumber
\end{align}
where $\mu_{\text{\ce{SrTiO3}}}$ and $g^{\text{bulk}}_{\text{\ce{SrTiO3}}}$ respectively refer to the chemical potential and the Gibbs free energy of the \ce{SrTiO3} unit cell.
We obtain:
\begin{align}
    \Omega^{\ce{SrTiO3}}_{\text{surf}} &= G_{\text{slab}}- N_{\text{Ti}}g^{\text{bulk}}_{\text{\ce{SrTiO3}}} - \Gamma^{\text{Ti}}_{\text{Sr}}\musr - \Gamma^{\text{Ti}}_{\text{O}}\muo
\end{align} where $\Gamma^{\text{Ti}}_{a} = N_{a} - N_{\text{Ti}}\frac{N^{\text{bulk}}_{a}}{N^{\text{bulk}}_{\text{Ti}}}$ refers to the excess $a$ component in the surface with respect to the number of Ti atoms and $\frac{N^{\text{bulk}}_a}{N^{\text{bulk}}_{\ce{Ti}}}$ refers to the bulk stoichiometric ratio of $a$ to Ti.

Approximating Gibbs free energies by DFT energies \cite{reuter_composition_2001, heifets_density_2007} and redefining chemical potentials by subtracting reference state energies $E_a$ for each component $a$ obtained from DFT calculations:
\begin{align}\label{equation:final-surf}
    \Omega^{\ce{SrTiO3}}_{\text{surf}} &= \phi - \Gamma^{\text{Ti}}_{\text{Sr}}\musr - \Gamma^{\text{Ti}}_{\text{O}}\muo
\end{align}
where 
\begin{align}
    \phi &\approx E_{\text{slab}}- N_{\text{Ti}}E^{\text{bulk}}_{\text{\ce{SrTiO3}}} - \Gamma^{\text{Ti}}_{\text{Sr}}E^{\text{bulk}}_{\text{\ce{Sr}}} - \Gamma^{\text{Ti}}_{\text{O}}\frac{E_{\text{\ce{O2}}}}{2} \nonumber
\end{align}

$E_{\text{slab}}$ is the slab energy, $E^{\text{bulk}}_{\text{\ce{SrTiO3}}}$ and $E^{\text{bulk}}_{\text{\ce{Sr}}}$ are the DFT energies of the \ce{SrTiO3} and \ce{Sr} bulk respectively. $E_{\text{\ce{O2}}}$ is the DFT energy of an isolated \ce{O2} molecule. In this work, we did not correct for oxygen overbinding in the gas phase. Even with such a correction, we only expect a minor change of the cross-over points and a downward shift in our surface phase diagram.

\subsection{Phase diagram creation}\label{methods:phase-diagram}
\ce{SrTiO3} has three elements but only two degrees of freedom in chemical potential as in Equation \ref{equation:final-surf}. The chemical potential of empty virtual sites is fixed. The MC-sampled structures exhibit different $\Omega^{\ce{SrTiO3}}_{\text{surf}}$ as $\musr$ and $\muo$ change. By selecting the most stable structure at each $\musr$ and $\muo$, a 2D phase diagram with $\musr$ and $\muo$ axes was obtained. 

\subsection{Density-functional theory calculations}\label{methods:dft}
Vienna \textit{ab initio} Simulation Package (VASP) v.6.2.1 \cite{kresse_efficient_1996} was employed to relax bulk structures and for single-point DFT calculations of \ce{SrTiO3}(001) surfaces, using the projector augmented-wave (PAW) method to describe core electrons \cite{kresse_ultrasoft_1999}. The Perdew-Burke-Ernzerhof (PBE) functional \cite{perdew_generalized_1996} within the generalized-gradient approximation (GGA) was utilized for spin-polarized calculations. Dipole corrections to the total energy were enabled along the z-axis for surfaces. The kinetic energy cutoff for plane waves was set to 520 eV. In the self-consistent field cycle, a limit of 10\textsuperscript{-6} eV was adopted as the stopping criterion.

\subsection{Workflow management and compute time}\label{methods:workflow}
An internal library, HTVS (for high-throughput virtual simulations) managed the DFT calculations, NFF training, and adversarial attacks. VSSR-MC and latent space clustering were run in separate procedures. DFT calculations took about 0.5-1 hour each on 30 cores of an Intel Xeon Platinum 8260 CPU. NFF training increased with dataset size and approached a maximum of 3 hours per model with an Nvidia Volta V100 32 GB GPU. VSSR-MC runs sampled at a rate of about an hour per $10,000$ iterations using the NFF energy model (${\sim}20,000$ iterations for each run for \sto{}) on an Nvidia GeForce RTX 2080 Ti 11 GB GPU and about 15 minutes per $10,000$ iterations using the classical Tersoff potential (${\sim}50,000$ iterations for each run for \gan{}) on 4 cores of an Intel Core i9-7920X CPU. For the \si{} 3x3 surface using the Stephenson, Radny \& Smith (SRS) modified Stillinger-Weber potential implemented in OpenKIM \cite{tadmor_potential_2011}, it took about 5 minutes per $10,000$ iterations (${\sim}50,000$ iterations for each run) on 4 cores of the same Intel Core i9-7920X CPU. Meanwhile, the \si{} 5x5 and 7x7 runs were carried out on the same machine used for DFT calculations (4 cores of an Intel Xeon Platinum 8260 CPU). The \si{} 5x5 surface required just 4 minutes per $10,000$ iterations (${\sim}150,000$ iterations for each run) and the \si{} 7x7 surface required 15 minutes per $10,000$ iterations (${\sim}200,000$ iterations for each run). Adversarial attack and latent space clustering were comparatively fast, taking less than 15 minutes for each generation on an Nvidia Volta V100 32 GB GPU and an Nvidia GeForce RTX 2080 Ti 11 GB GPU respectively.

\section{Data availability}
The trained models, DFT data, and Jupyter notebooks used for data analysis are available on Zenodo: \url{https://doi.org/10.5281/zenodo.7758174} \cite{du_data_2023}.

\section{Code availability} 
The VSSR-MC algorithm reported in this work is available on GitHub: \url{https://github.com/learningmatter-mit/surface-sampling}. The version of code used in this work is available on Zenodo: \url{https://doi.org/10.5281/zenodo.10086398} \cite{du_2023_code}.

\section{Acknowledgements}
The authors thank Gavin Winter, Jiayu Peng, Nathan Frey, and Mengren Liu for helpful discussions. The authors also appreciate manuscript editing by Jiayu Peng and Alexander Hoffman. X.D. acknowledges support from the National Science Foundation Graduate Research Fellowship under Grant No. 2141064. J.K.D. was supported by the Department of Defense through the National Defense Science \& Engineering Graduate Fellowship Program. The authors are grateful for computation time allocated on the MIT SuperCloud cluster, the MIT Engaging cluster, and the NERSC Perlmutter cluster. This material is based upon work supported by the Under Secretary of Defense for Research and Engineering under Air Force Contract No. FA8702-15-D-0001. Any opinions, findings, conclusions or recommendations expressed in this material are those of the author(s) and do not necessarily reflect the views of the Under Secretary of Defense for Research and Engineering. © 2023 Massachusetts Institute of Technology. Delivered to the U.S. Government with Unlimited Rights, as defined in DFARS Part 252.227-7013 or 7014 (Feb 2014). Notwithstanding any copyright notice, U.S. Government rights in this work are defined by DFARS 252.227-7013 or DFARS 252.227-7014 as detailed above. Use of this work other than as specifically authorized by the U.S. Government may violate any copyrights that exist in this work.

\section{Author contributions statement}
X.D. implemented the sampling algorithm, performed surface modeling, ran DFT calculations, trained the neural networks, and carried out surface stability analysis.
J.K.D. assisted with sampling algorithm implementation and provided guidance with surface modeling. 
J.R.L. provided guidance with surface modeling and ran DFT calculations.
R.M. provided guidance with neural network training and active learning.
B.Y. provided guidance with the choice of surfaces and surface stability analysis.
L.L. supervised the research and contributed to securing funding.
R.G.-B. conceived the project, supervised the research, and contributed to securing funding.
All contributed to results discussion and manuscript writing.

\section{Competing interests statement}
The authors declare no competing financial or non-financial interests.

\newpage
\captionsetup[figure]{labelfont={bf},name={Fig.},labelsep=colon}

\section{Figures}

\begin{figure}[htb!]%
\centering
\includegraphics[width=1.0\textwidth]{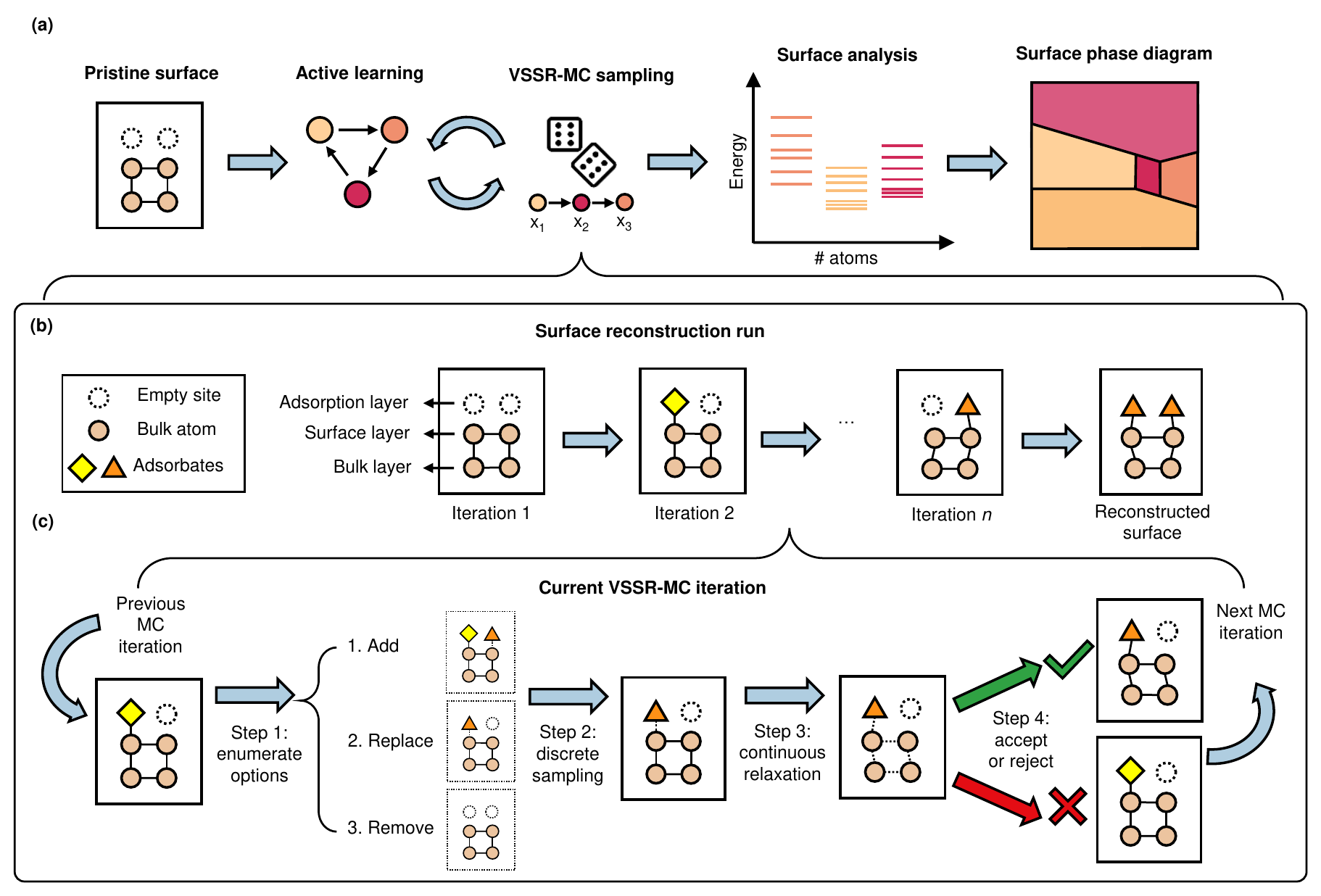}
\caption{\textbf{Automatic Surface Reconstruction framework.} (a) Beginning with a pristine surface and computer-generated virtual adsorption sites, VSSR-MC sampling is conducted in tandem with active learning of an NFF. The Monte Carlo nature of VSSR-MC is denoted by a pair of dice. Following multiple rounds of surface reconstruction runs, the surface free energy of sampled structures are separated by their composition and analyzed at different external conditions. Finally, the surface phase diagram is constructed, with each color representing the dominant surface across a set of external conditions. (b) The proposed surface reconstruction run consists of multiple iterations of VSSR-MC. Starting from the same pristine surface with empty virtual sites, a series of iterations alter both the atomic identities at adsorption sites and the relative atomic positions within the structure. (c) In each VSSR-MC iteration, there are three steps. Starting from a surface taken from the previous iteration, in step 1, possible options for discrete sampling on a single adsorption site are enumerated. The possible actions are: add, replace, and remove. In step 2, the chosen discrete sampling action is performed to produce a candidate structure. This mutation is followed by continuous relaxation in step 3. For step 3, an energy model is required, such as an NFF or a classical force field. Finally, in step 4, the candidate structure is evaluated using an MC acceptance criterion.}\label{fig1}
\end{figure}

\begin{figure}[htb!]%
\centering
\includegraphics[width=0.6\textwidth]{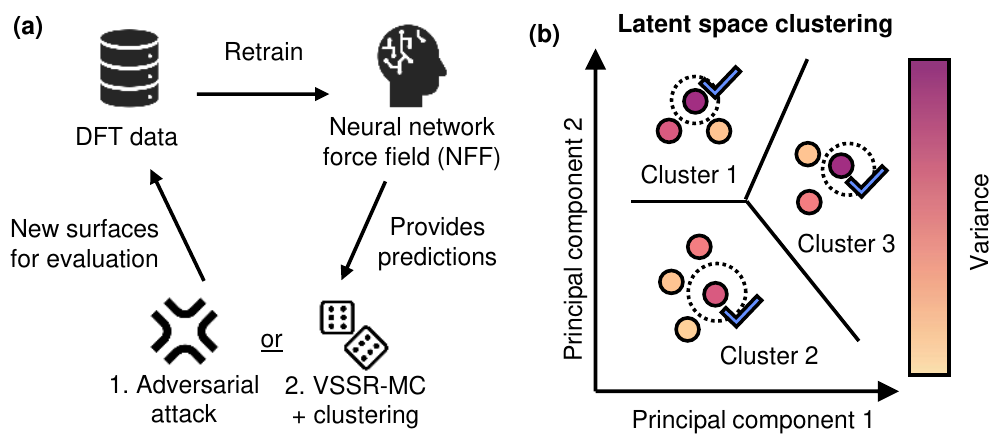}
\caption{\textbf{Active learning procedure for neural network force field.} (a) An initial ensemble of NFF models is trained using a common dataset of available DFT data. Using the NFF to provide predictions of forces, energies and their SD, either adversarial attack or VSSR-MC with latent space clustering is used to generate new surfaces for DFT evaluation. After performing DFT calculations on these new structures, they are added to the data pool and the NFF is retrained. (b) Latent space clustering procedure for MC-generated structures. Structures are clustered according to the first few principal components of latent space embeddings and the structure with the highest force SD in each cluster is selected for DFT evaluation.}\label{fig2}
\end{figure}

\begin{figure}[htb!]%
\centering
\includegraphics[width=1.0\textwidth]{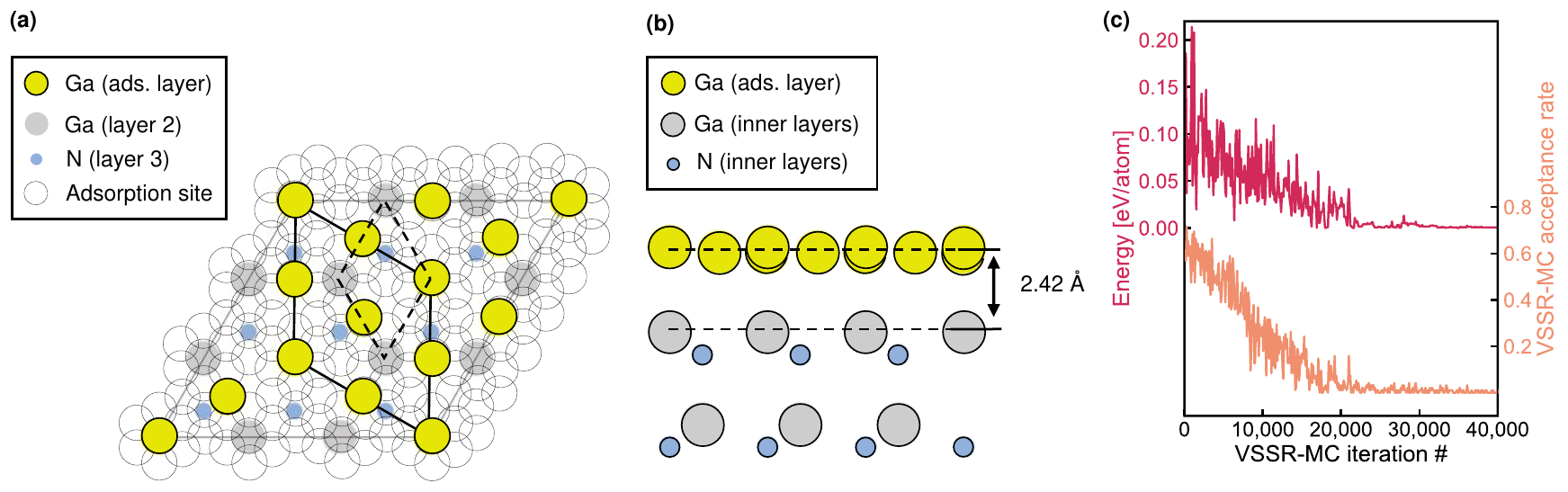}
\caption{\textbf{Obtaining the \gan{} contracted-monolayer surface reconstruction using VSSR-MC and a classical potential.} (a) Top view of surface reconstruction. Only the first three layers are shown. ``Ads. layer" refers to the adsorption layer. Our structure is visually identical to the literature optimized structure from Northrup \textit{et al.} [46] with an energy difference of 0.008 meV/atom. (b) Side view of reconstructed surface. The average distance of the first two layers is 2.42 Å, in agreement with literature. The dashed lines are a guide for the eye. (c) Typical VSSR-MC run profile for \gan{}. Annealing from a high temperature allows convergence to a low-energy structure.}\label{fig3}
\end{figure}

\begin{figure}[htb!]%
\centering
\includegraphics[width=0.7\textwidth]{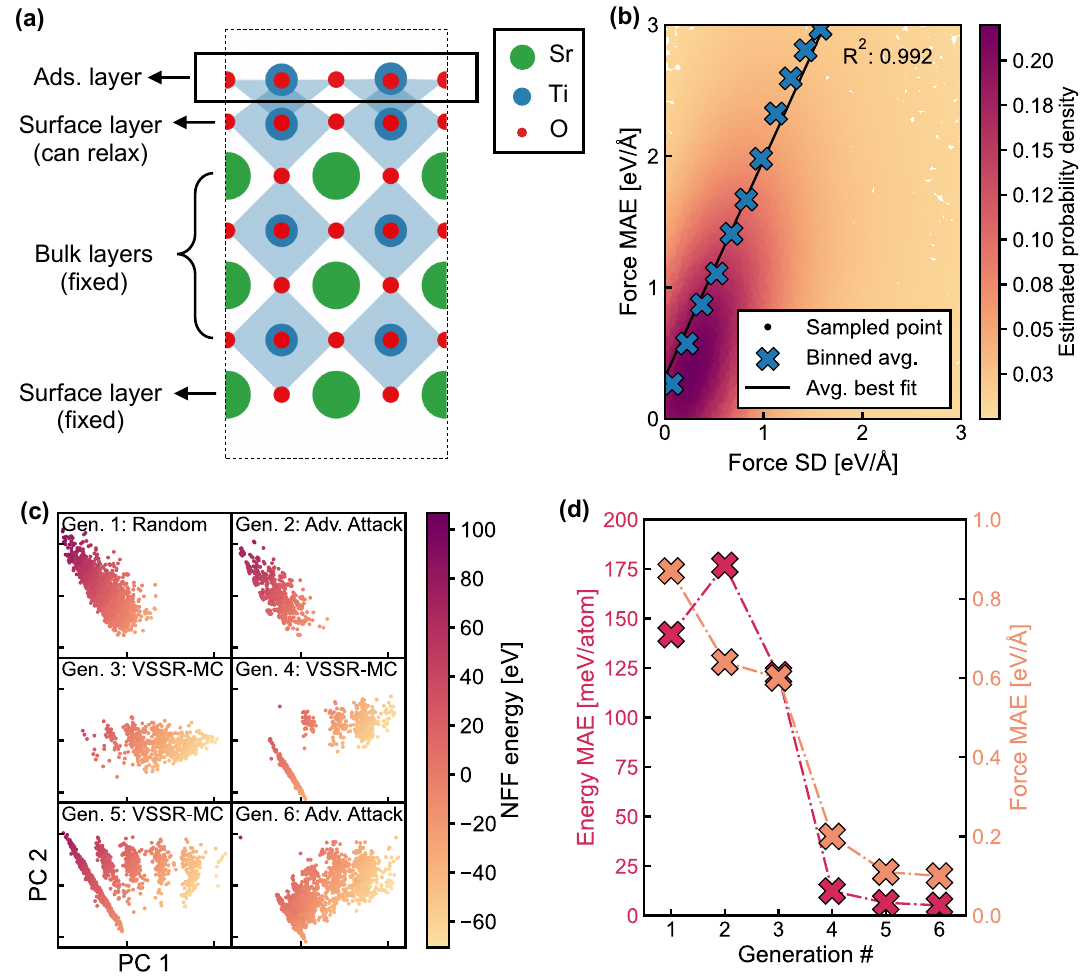}
\caption{\textbf{Active learning for neural network force field.} (a) Example \sto{} surface for training NFF. The surface consists of an adsorption (ads.) layer with variable chemical compositions and configurations, and inner layers with fixed compositions. The blue diamonds represent the octahedral of O atoms surrounding a Ti atom. (b) Correlation plot of force MAE with force SD. The binned average is calculated by dividing both the force SD and force MAE into equal-sized bins. The average force MAE is then plotted against the median force SD for each corresponding bin. The average best fit line is based on the binned average values and the trend implies force SD is a fair surrogate for estimating force MAE. (c) Principal component analysis (PCA) of surfaces obtained at each AL generation. Starting with generation 2, a mix of adversarial attack and VSSR-MC with latent space clustering were used. (d) Retrospective performance of NFF trained using all surfaces available at each AL generation. Performance was measured using a test set consisting of VSSR-MC structures sampled from the 6th generation NFF model to reflect our use case.
}\label{fig4}
\end{figure}

\begin{figure}[htb!]%
\centering
\includegraphics[width=1.0\textwidth]{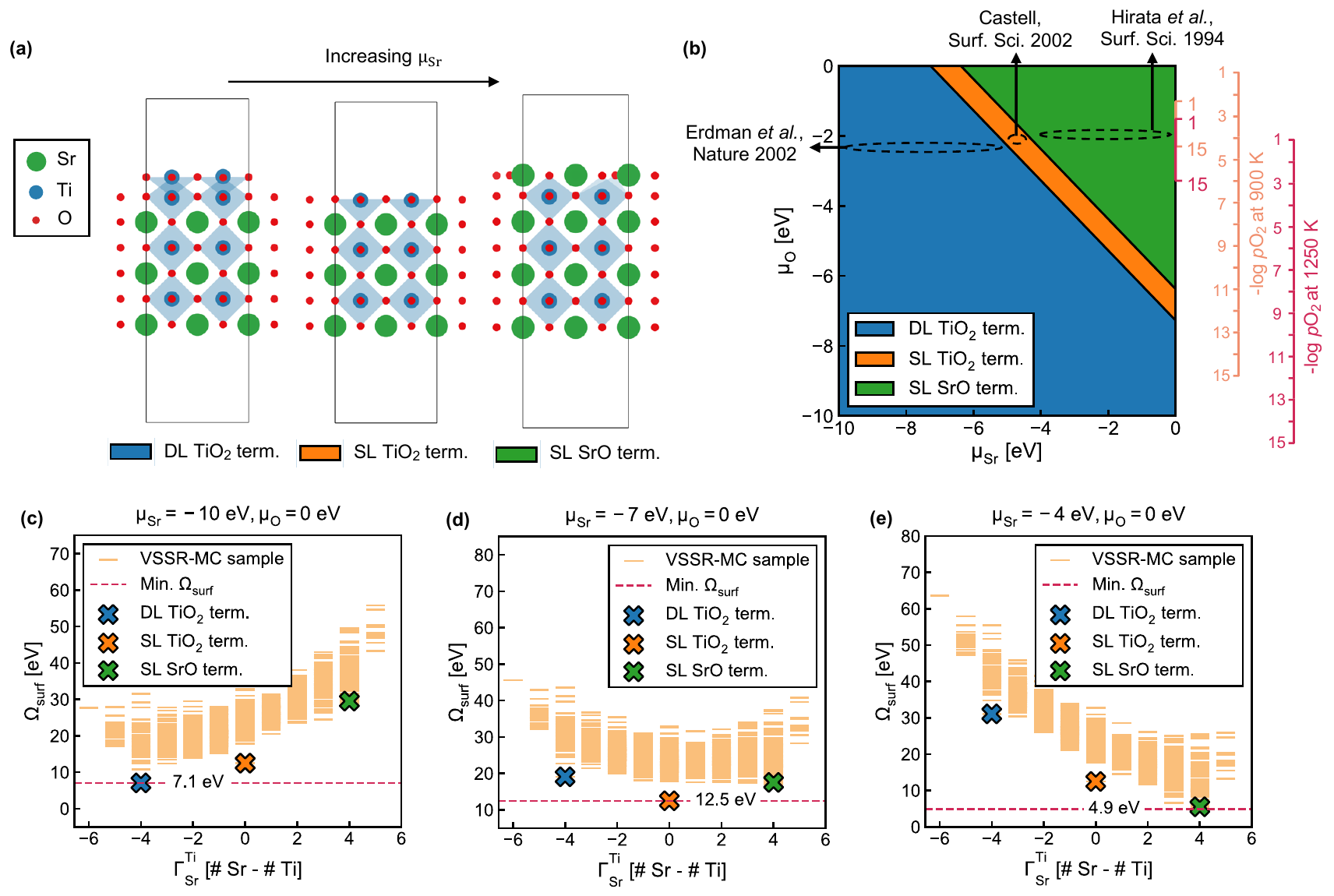}
\caption{\textbf{Analyzing dominant \sto{} surface terminations observed during VSSR-MC runs at different chemical potentials.} (a) Side-view of change in dominant surface termination as $\musr$ increases. The blue diamonds represent the octahedral of O atoms surrounding a Ti atom. (b) Computed phase diagram showing the stable surface terminations at varying $\musr$ and $\muo$ along with estimated positions of three experimental \sto{} surfaces, Erdman \textit{et al.} [11], Castell [10], and Hirata \textit{et al.} [9]. On the right, four vertical axes are illustrated. The smaller axes on the phase diagram, ending at points 1 and 15, provide an abbreviated view. The larger scales extend from 1 to 15 at equal intervals, indicating $-\log p_{\ce{O2}}$ values at 900 K (orange) and 1250 K (red), offering a detailed perspective on the oxygen partial pressure across these temperatures. (c-e) Surface free energy ($\Omega_{\text{surf}}$) plots of sampled structures as a function of the difference in the number of Sr and Ti atoms. Plots shown correspond to $\Omega_{\text{surf}}$ at various $\musr$: (c) $\musr = -10$ eV, (d) $\musr = -7$ eV, and (e) $\musr = -4$ eV. $\muo = 0$ eV in all three plots. The minimum energy surfaces are crossed out and correspond to those from literature.}\label{fig5}
\end{figure}

\begin{figure}[htb!]%
\centering
\includegraphics[width=0.7\textwidth]{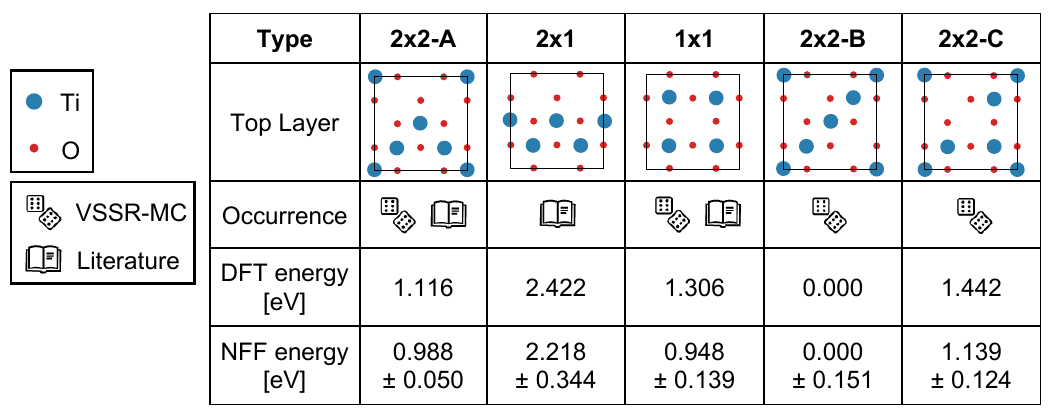}
\caption{\textbf{Comparing DFT and NFF energies of double-layer \ce{TiO2} terminations, all with the same composition.} The NFF predicted energies are close to DFT energies. Additionally, the previously-unreported 2x2-B and 2x2-C terminations have roughly equal or lower energies than those of the other three literature-reported surfaces.
}\label{fig6}
\end{figure}
\FloatBarrier

\newpage
\captionsetup[figure]{labelfont={bf},name={Extended Data Fig.},labelsep=colon}
\captionsetup[table]{labelfont={bf},name={Extended Data Table},labelsep=colon}
\setcounter{figure}{0}

\section{Extended Data Figures}
\begin{figure}[htb!]%
\centering
\includegraphics[width=0.8\textwidth]{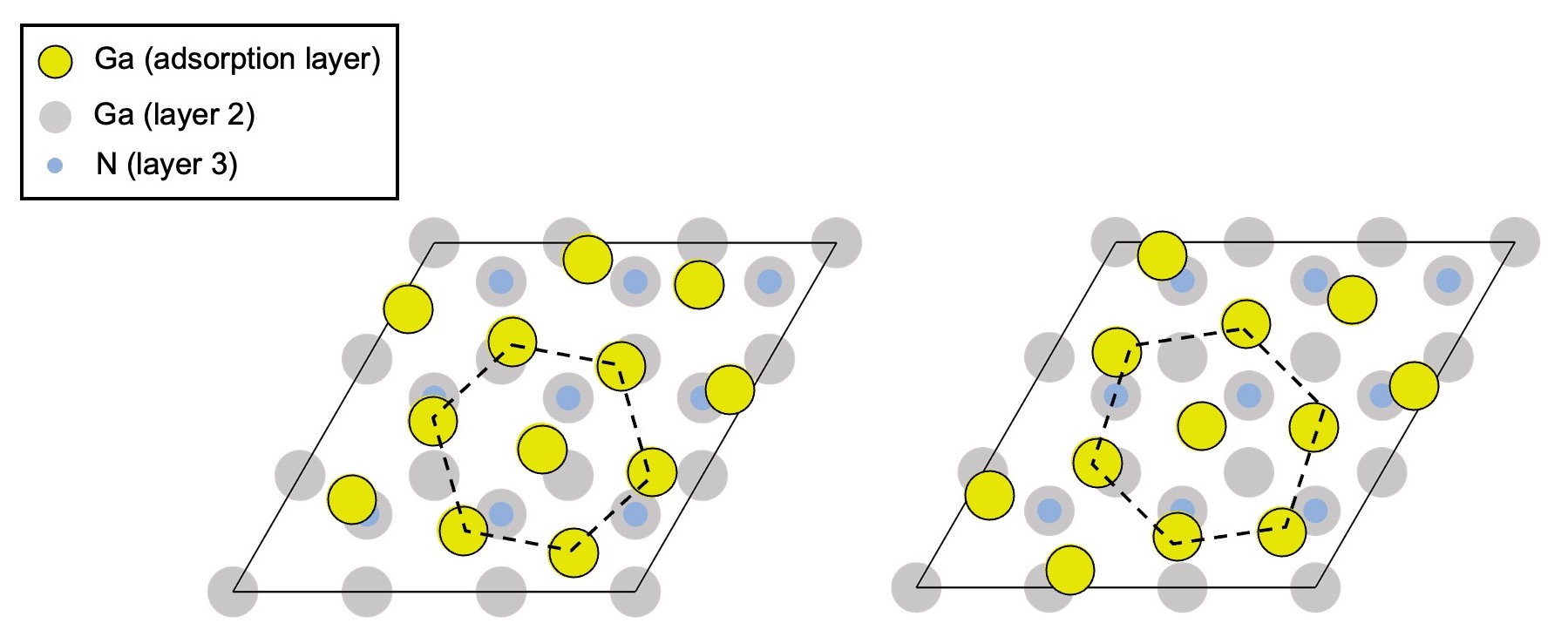}
\caption{\textbf{Top view of additional GaN(0001) MC-sampled structures.} The surface reconstructions are rotated in comparison with the reference structure from Northrup \textit{et al.} [46] but contain the same hexagonal pattern.}\label{EDfig1}
\end{figure}

\begin{figure}[htb!]%
\centering
\includegraphics[width=0.85\textwidth]{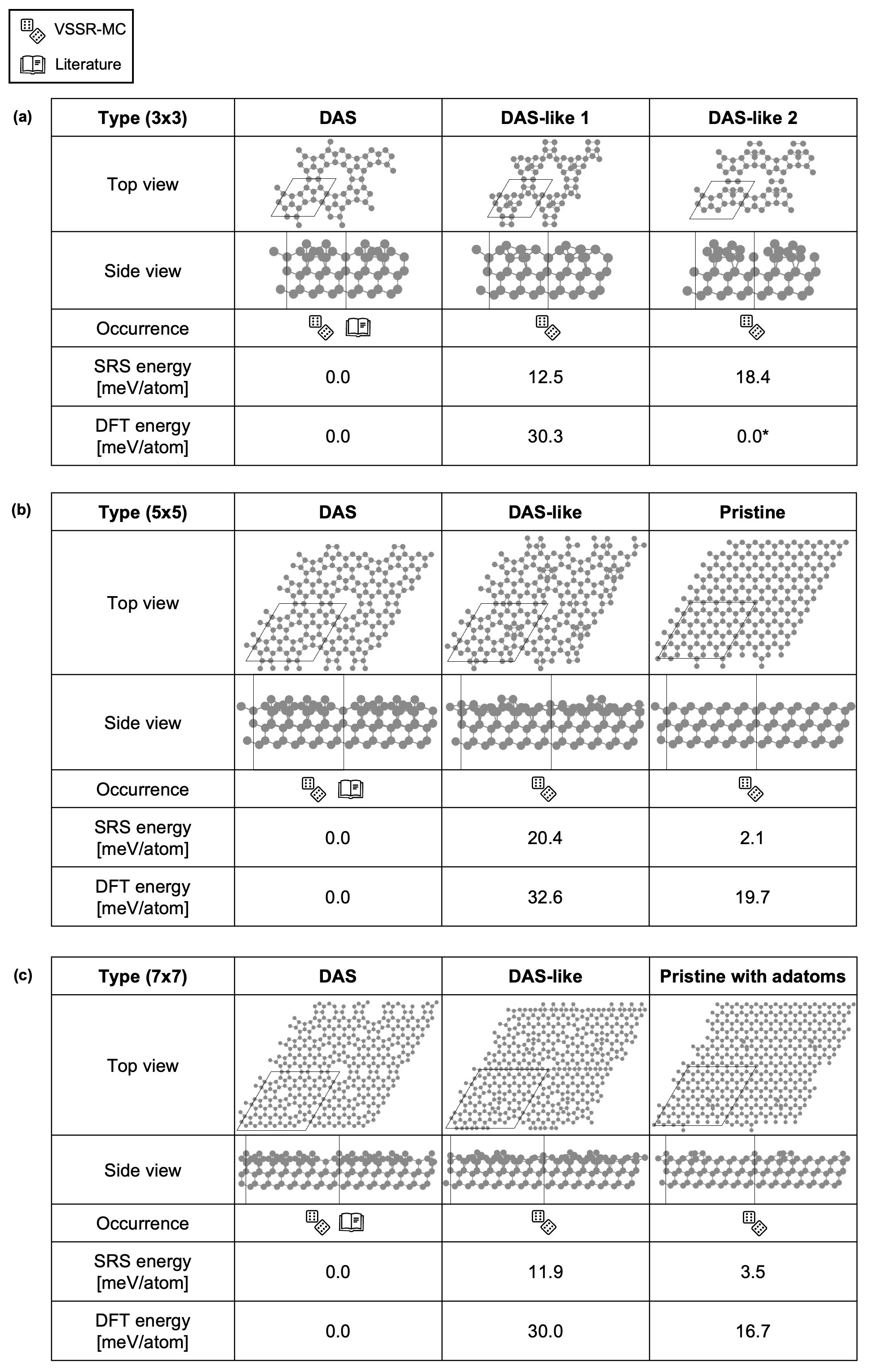}
\caption{\textbf{Comparing classical potential and DFT energies of Si(111) sampled surface reconstructions.} Structures shown were obtained from constant-composition (canonical) VSSR-MC sampling using the SRS modified Stillinger-Weber potential [45] with (a) 3x3, (b) 5x5, and (c) 7x7 unit cells. The SRS energies were obtained from the depicted structures while the DFT energies came from structures further relaxed at the DFT level. * Further relaxation using DFT resulted in the 3x3 DAS structure.}\label{EDfig2}
\end{figure}

\begin{figure}[htb!]%
\centering
\includegraphics[width=0.5\textwidth]{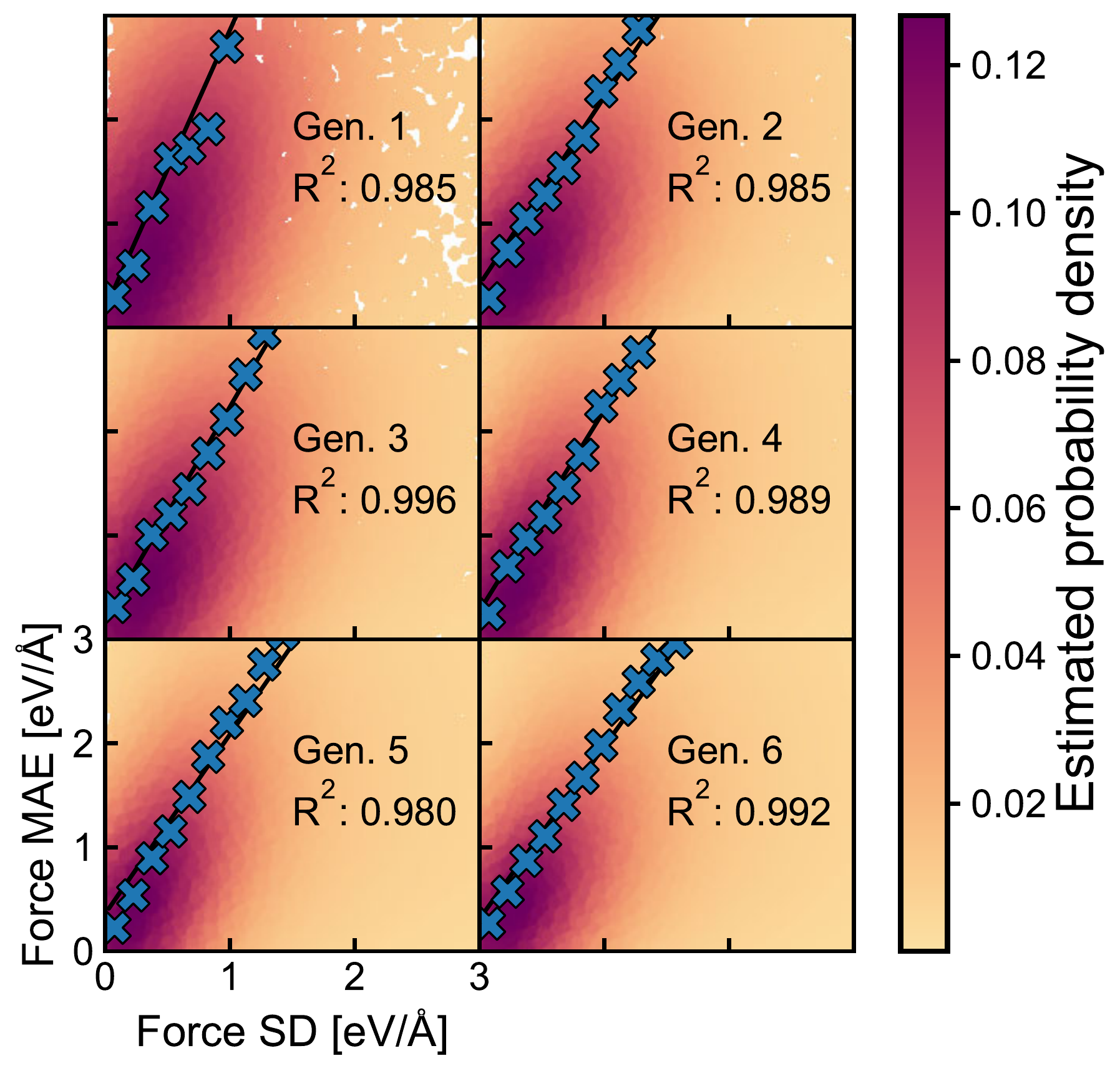}
\caption{\textbf{Correlation plot of force MAE with force SD over AL generations.} At each AL generation, an ensemble of just three NFF models was able to estimate force SD that correlated strongly with force error. Each individual data point represents a sampled structure. Each blue ‘X’ represents a binned average, and a best fit line is drawn through the binned averages. The binned average is calculated by dividing both the force SD and force MAE into equal-sized bins. The average force MAE is then plotted against the median force SD for each corresponding bin.}\label{EDfig3}
\end{figure}

\begin{figure}[htb!]%
\centering
\includegraphics[width=0.5\textwidth]{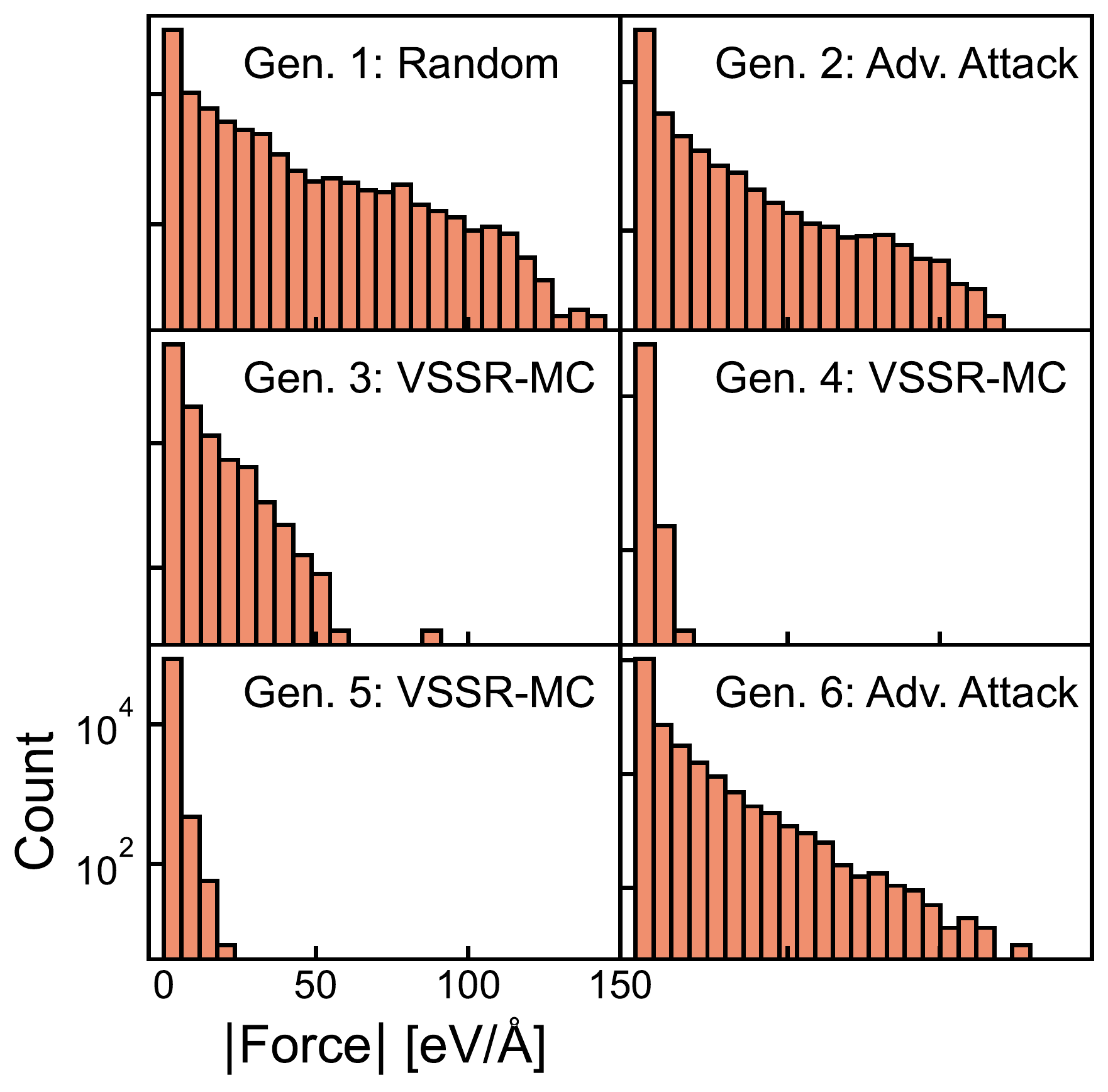}
\caption{\textbf{Force distribution over AL generations.} The majority of high-force structures were added in AL generations 1, 2, and 6, which correspond either to random structures or structures obtained through adversarial attack. The three VSSR-MC AL generations produced structures with low force values mostly around 50 eV/Å or less.}\label{EDfig4}
\end{figure}

\begin{figure}[htb!]%
\centering
\includegraphics[width=1.0\textwidth]{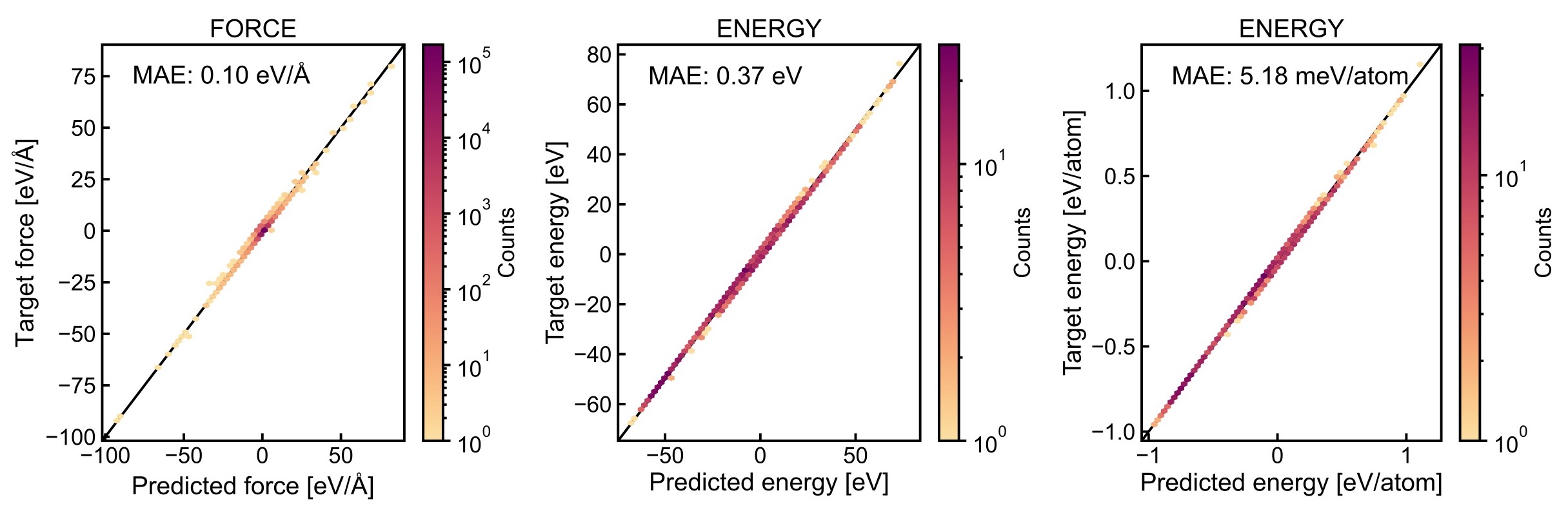}
\caption{\textbf{Test performance of the best NFF model.} As described in the main paper, the test data is obtained from VSSR-MC runs using the 6th generation NFF model.}\label{EDfig5}
\end{figure}

\begin{figure}[htb!]%
\centering
\includegraphics[width=1.0\textwidth]{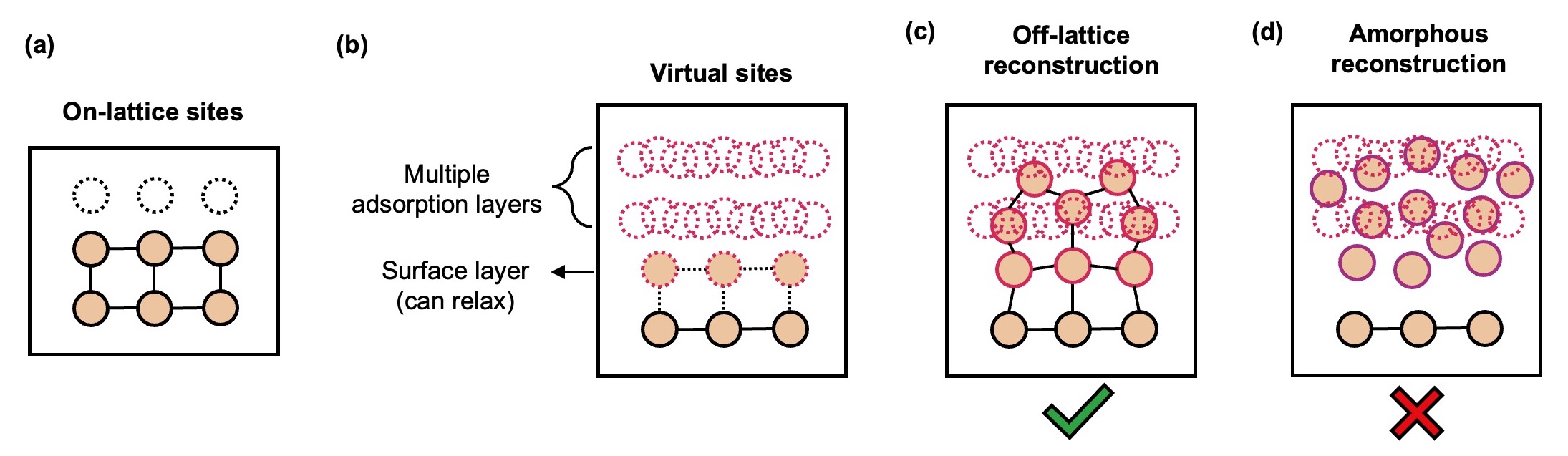}
\caption{\textbf{Strengths and limitations of VSSR-MC.} Comparison of (a) limited fixed on-lattice sites and (b) denser algorithmically-generated virtual surface sites that can overlap. (c) Off-lattice reconstructions can be obtained following VSSR-MC discrete sampling at virtual sites and continuous relaxation of surface atoms and adsorbates. (d) Amorphous reconstructions with many local minima, however, will likely be difficult for VSSR-MC to sample.
}\label{EDfig6}
\end{figure}

\begin{figure}[htb!]%
\centering
\includegraphics[width=0.7\textwidth]{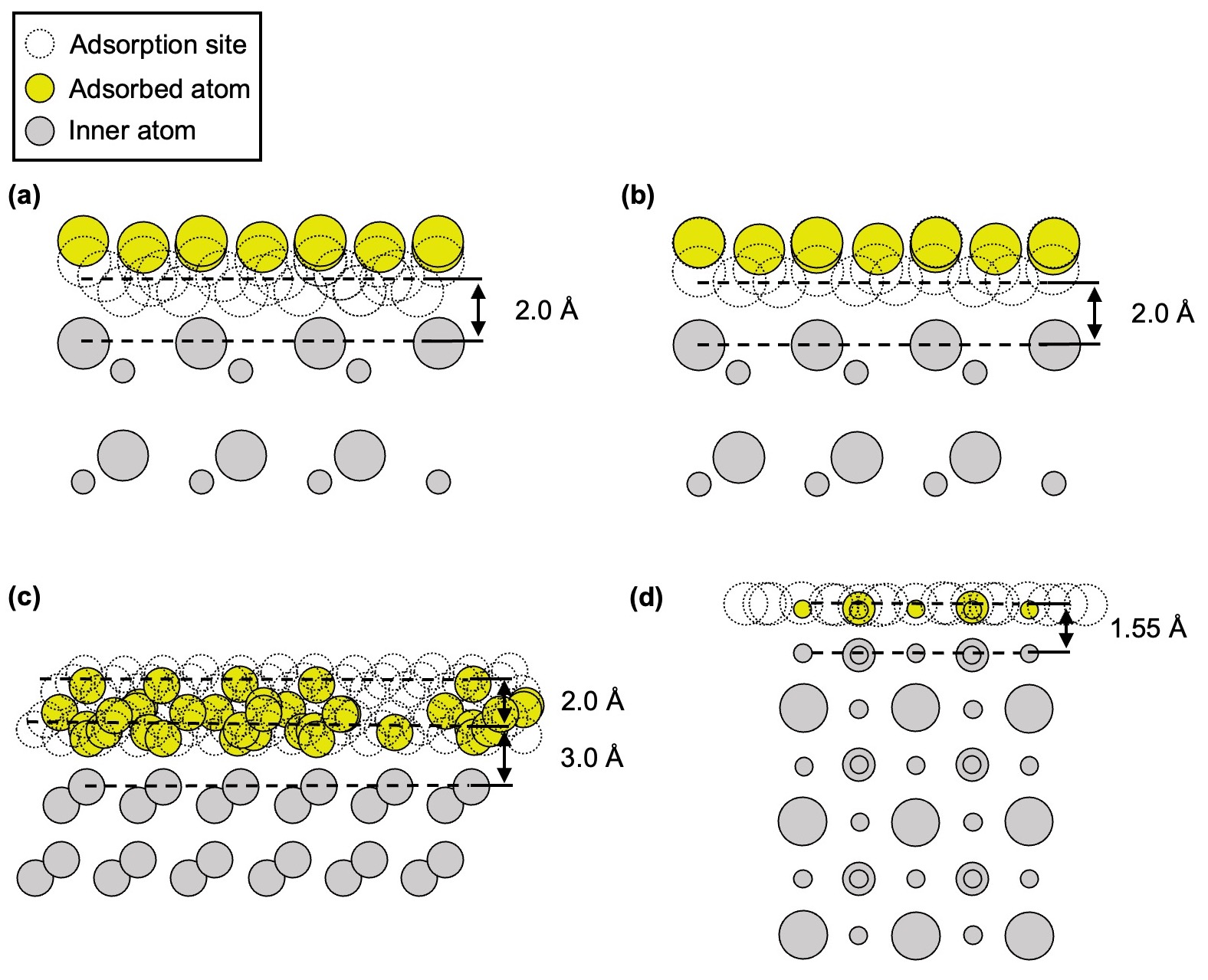}
\caption{\textbf{Side view of virtual sites for surfaces studied in this work.} (a) pymatgen and (b) CatKit virtual sites for GaN(0001) against the contracted Ga monolayer reconstruction, (c) two-layer pymatgen sites for Si(111) against the 5x5 DAS reconstruction, and (d) pymatgen virtual sites for \ce{SrTiO3}(001) against the DL \ce{TiO2} reconstruction. The dashed lines are a guide for the eye.
}\label{EDfig7}
\end{figure}

\begin{figure}[htb!]%
\centering
\includegraphics[width=0.7\textwidth]{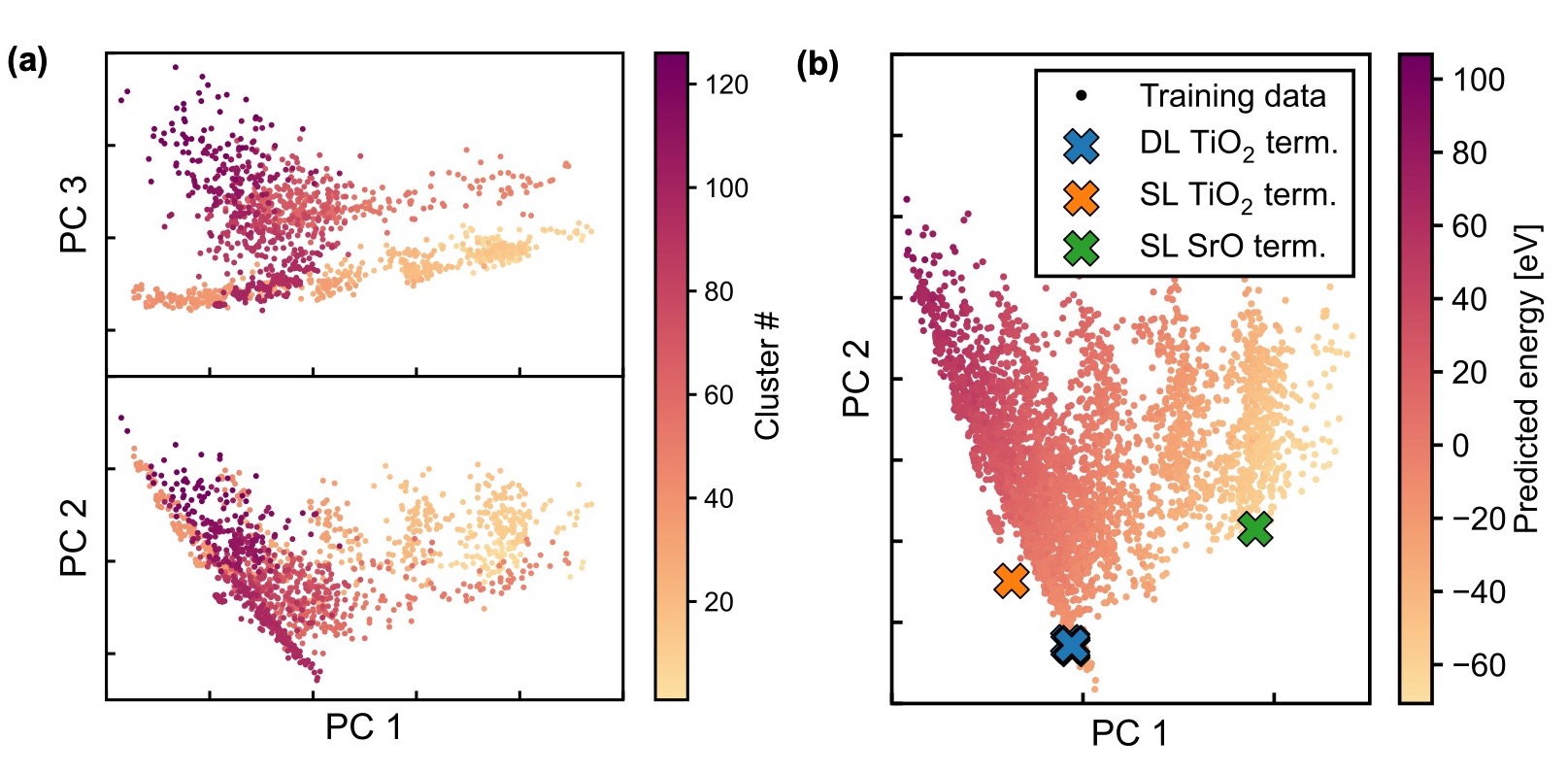}
\caption{\textbf{Visualizations in the latent space.} (a) Clustering of VSSR-MC structures in the NFF latent space visualized in the first three principal components. In the VSSR-MC with clustering AL method, the surface from each cluster with the highest force SD is selected for DFT evaluation. (b) PCA of training data and the dominant terminations (term.) in the latent space of the 6th generation model.
}\label{EDfig8}
\end{figure}

\clearpage
\section{Extended Data Table}

\vspace*{\fill}
\begin{center}
     Table on the next page.
\end{center}
\vspace*{\fill}

\begin{sidewaystable}[htb!]
    \captionsetup{width=1.0\textwidth}
    \centering
    \includegraphics[width=1.0
    \textwidth]{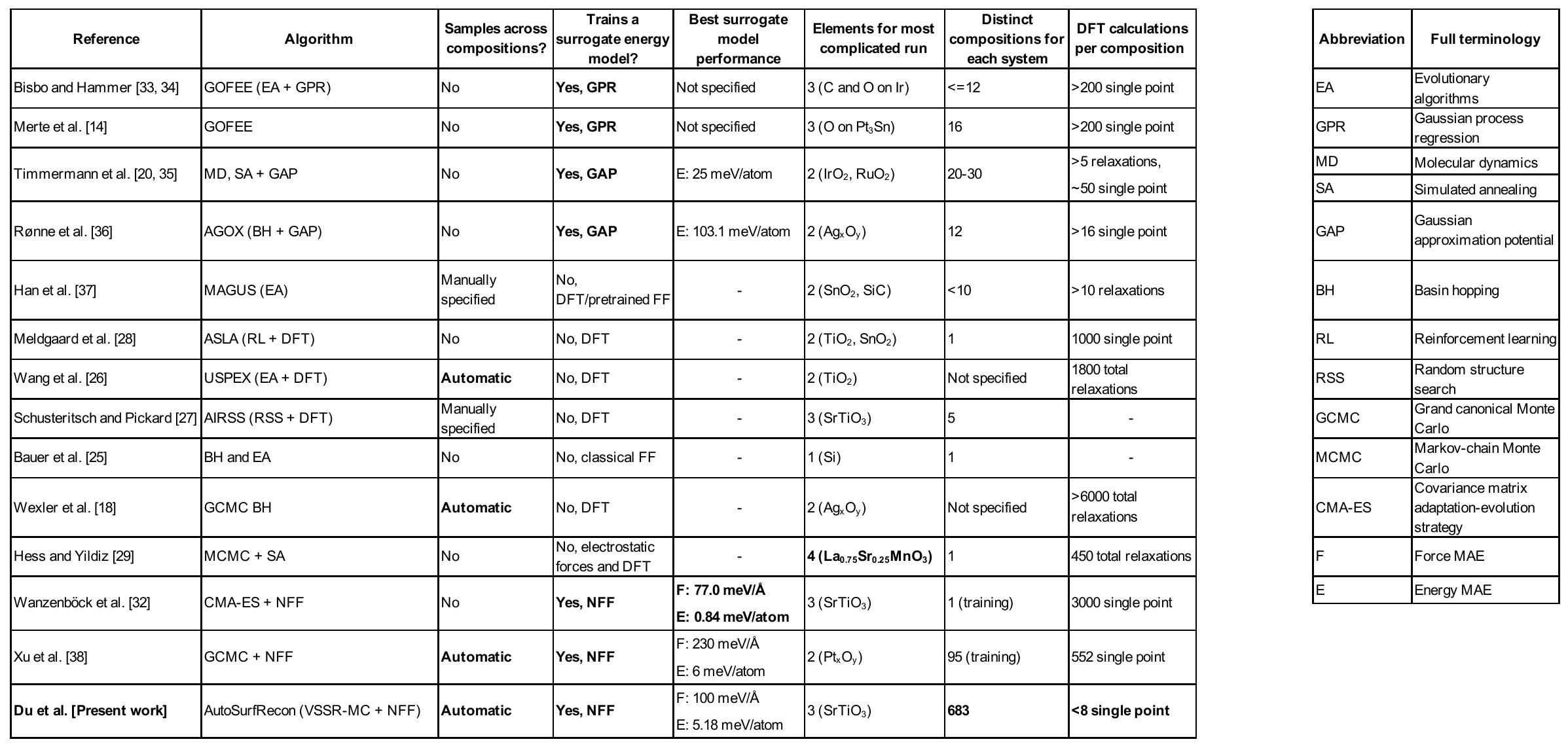}
    \caption[width=0.7]{\textbf{Comparison of AutoSurfRecon with existing computational methods for surface reconstruction.} AutoSurfRecon automatically samples across many surface compositions and configurations while training an accurate NFF for low-cost energy prediction.}
    \label{EDtab1}
\end{sidewaystable}
\FloatBarrier

\newpage
\setcounter{equation}{0}
\setcounter{section}{0}
\setcounter{page}{1}
% set equations to begin with S
\renewcommand{\theequation}{S\arabic{equation}}
% set page to begin with S
\renewcommand{\thepage}{S\arabic{page}}
% set sections to begin with Supplementary Section
\renewcommand{\thesection}{Supplementary Section \arabic{section}} 

\begin{center}
\Large
Supplementary information for ``Machine-learning-accelerated simulations to enable automatic surface reconstruction"
\break

\normalsize
Xiaochen Du$^{1,2}$, James K. Damewood$^{2,3}$, Jaclyn R. Lunger$^{3}$, Reisel Millan$^{3}$, Bilge Yildiz$^{3,4,5}$, Lin Li$^{6}$, and Rafael Gómez-Bombarelli$^{3,*}$
\break

${}^1$\ChemE, \MIT, \MITaddress\\
${}^2$\CSE, \MIT, \MITaddress\\
${}^3$\DMSE, \MIT, \MITaddress\\
${}^4$\Nuclear, \MIT, \MITaddress\\
${}^5$\MTL, \MIT, \MITaddress\\
${}^6$\LL, \LLaddress\\
${}^*$Corresponding author: Rafael Gómez-Bombarelli, rafagb@mit.edu\\
\end{center}
  
% \author[1,2]{Xiaochen Du}
% \author[2,3]{}
% \author[3]{}
% \author[3]{}
% \author[3,4,5]{}
% \author[6]{}
% \author[3,*]{}
% \affil[1]{}
% \affil[2]{}
% \affil[3]{}
% \affil[4]{}
% \affil[5]{}
% \affil[6]{}
% \affil[*]{}

\break
\section{Abbreviations used}

The following is the list of abbreviations utilized.
\begin{itemize}
    \setlength\itemsep{0em}
    \item DFT: Density-functional theory
    \item MC: Monte Carlo
    \item MCMC: Markov-chain Monte Carlo
    \item ML: Machine learning
    \item NFF: Neural network force field
    \item AL: Active learning
    \item AutoSurfRecon: Automatic Surface Reconstruction
    \item VSSR-MC: Virtual Surface Site Relaxation-Monte Carlo
    \item DAS: Dimer-adatom stacking fault
    \item SD: Standard deviation
    \item MAE: Mean absolute error
    \item PCA: Principal component analysis
    \item DL: Double layer
    \item SL: Single layer
    \item LR: Learning rate
    \item SRS: Stephenson, Radny \& Smith (potential)
\end{itemize}

\clearpage
\section{Surface stability analysis}
The stability of surfaces can be compared using the surface Gibbs free energy $\Omega_{\text{surf}}$ \cite{reuter_composition_2001, heifets_density_2007, heifets_electronic_2007}:
\begin{align}
    \Omega_{\text{surf}} &= G_{\text{slab}} - \sum_{\text{all } a} N_{a}\mu_a
\end{align}
where $G_{\text{slab}}$ refers to the Gibbs free energy of the slab. For each element $a$, $N_{a}$ refers to the number of $a$ atoms in the slab, $\mu_a$ refers to the chemical potential of $a$.

For \ce{SrTiO3} specifically, we have its surface Gibbs free energy:
\begin{align}
    \Omega^{\ce{SrTiO3}}_{\text{surf}} &= G_{\text{slab}} - N_{\text{Sr}}\musr - N_{\text{Ti}}\muti - N_{\text{O}}\muo
\end{align}

The \ce{SrTiO3} chemical potential is defined as the sum of the chemical potentials of the component elements:
\begin{align}
    \mu_{\text{\ce{SrTiO3}}} &= \musr + \muti + 3\muo \nonumber
\end{align}

At thermodynamic equilibrium, the chemical potential of the surface is equal to the bulk crystal Gibbs free energy:
\begin{align}
    \mu_{\text{\ce{SrTiO3}}} &= g^{\text{bulk}}_{\text{\ce{SrTiO3}}} \nonumber
\end{align}

Substituting, we obtain:
\begin{align}
    \Omega^{\ce{SrTiO3}}_{\text{surf}} &= G_{\text{slab}}- N_{\text{Ti}}g^{\text{bulk}}_{\text{\ce{SrTiO3}}} - \Gamma^{\text{Ti}}_{\text{Sr}}\musr - \Gamma^{\text{Ti}}_{\text{O}}\muo
\end{align} where $\Gamma^{\text{Ti}}_{a} = N_{a} - N_{\text{Ti}}\frac{N^{\text{bulk}}_{a}}{N^{\text{bulk}}_{\text{Ti}}}$ refers to the excess $a$ component in the surface with respect to the number of Ti atoms and $\frac{N^{\text{bulk}}_a}{N^{\text{bulk}}_{\ce{Ti}}}$ refers to the bulk stoichiometric ratio of $a$ to Ti.

The bulk Gibbs free energy for a crystal can be decomposed into the following:
\begin{align}
    g^{\text{bulk}}_{a} &= E^{\text{bulk}}_{a} + E^{\text{vib}}_a - Ts_a + pv_a
\end{align} where $E^{\text{bulk}}_{a}$ is the static crystal energy obtained through DFT calculations, $E^{\text{vib}}_a$ is the vibrational component, $T$ is the temperature, $s_a$ is the entropy of the crystal, $p$ is the pressure, and $v_a$ is the volume of the crystal. By performing approximations similar to \cite{reuter_composition_2001, heifets_density_2007}, changes in $E^{\text{vib}}_a$ due to temperature roughly cancel out changes in $Ts_a$ while $pv_a$ is negligible. 

Thus, Gibbs free energies can be approximated by the DFT energies:
\begin{align}
    g^{\text{bulk}}_{a} &\approx E^{\text{bulk}}_{a} \nonumber
\end{align}

For later convenience in plotting the phase diagram, bulk-subtracted chemical potentials are introduced for Sr and Ti:
\begin{align}
    \musr' &= \musr -  g^{\text{bulk}}_{\text{Sr}} \approx \musr -  E^{\text{bulk}}_{\text{Sr}}\\
    \muti' &= \muti -  g^{\text{bulk}}_{\text{Ti}} \approx \muti -  E^{\text{bulk}}_{\text{Ti}}
\end{align} where $E^{\text{bulk}}_{\text{Sr}}$ and $E^{\text{bulk}}_{\text{Ti}}$ are obtained from DFT calculations.

Similarly, $\muo'$ is defined with the reference state as an isolated oxygen molecule:
\begin{align}
    \muo' &= \muo - \frac{1}{2}E_{\ce{O2}}
\end{align} where $E_{\text{\ce{O2}}}$ is the DFT total energy of a single oxygen molecule.

Additionally, $\muo$ can be defined as a function of experimental conditions, \ce{O2} partial pressure ($p_{\ce{O2}}$) and temperature ($T$), with the following equation:
\begin{align}
    \muo(T,p_{\ce{O2}}) = \frac{1}{2}\left[ E_{\ce{O2}} + \mu_{\ce{O2}}(T, p_0) + k_B T\ln{\left(\frac{p_{\ce{O2}}}{p_0}\right)}\right]
\end{align} where $p_0$ = 1 bar is the reference pressure and $\mu_{\ce{O2}}(T, p_0)$ is the reference chemical potential obtained from NIST \cite{thomas_c_allison_nist-janaf_2013}.

We swap out $\mu_a$ for $\mu_a'$ and transform to obtain:
\begin{align}
    \Omega^{\ce{SrTiO3}}_{\text{surf}} &= \phi - \Gamma^{\text{Ti}}_{\text{Sr}}\musr' - \Gamma^{\text{Ti}}_{\text{O}}\muo'
\end{align}
where 
\begin{align}
    \phi &= G_{\text{slab}}- N_{\text{Ti}}g^{\text{bulk}}_{\text{\ce{SrTiO3}}} - \Gamma^{\text{Ti}}_{\text{Sr}}g^{\text{bulk}}_{\text{\ce{Sr}}} - \Gamma^{\text{Ti}}_{\text{O}}\frac{E_{\text{\ce{O2}}}}{2} \nonumber\\
    &\approx E_{\text{slab}}- N_{\text{Ti}}E^{\text{bulk}}_{\text{\ce{SrTiO3}}} - \Gamma^{\text{Ti}}_{\text{Sr}}E^{\text{bulk}}_{\text{\ce{Sr}}} - \Gamma^{\text{Ti}}_{\text{O}}\frac{E_{\text{\ce{O2}}}}{2} \nonumber
\end{align}
$G_{\text{slab}}$ is approximated by the slab energy $E_{\text{slab}}$. For thermodynamic stability, $\musr' < 0$, $\muo' < 0$.

Note: In our main text, we denote $\musr'$ and $\muo'$ as $\musr$ and $\muo$ respectively for simplicity.

VSSR-MC samples according to the grand potential $\Omega_{\text{G}} $, which is equivalent to $\Omega_{\text{surf}}$ after approximations:
\begin{align}
    \Omega^{\ce{SrTiO3}}_{\text{G}} &= E_{\text{slab}} - Ts_{\text{slab}} - N_{\text{Sr}}\musr - N_{\text{Ti}}\muti - N_{\text{O}}\muo \\
    &= G_{\text{slab}} - pv_{\text{slab}} - N_{\text{Sr}}\musr - N_{\text{Ti}}\muti - N_{\text{O}}\muo \nonumber\\
    &\approx G_{\text{slab}} - N_{\text{Sr}}\musr - N_{\text{Ti}}\muti - N_{\text{O}}\muo \nonumber\\
    &= \Omega^{\ce{SrTiO3}}_{\text{surf}} \nonumber
\end{align}


\begin{thebibliography}{10}
\expandafter\ifx\csname url\endcsname\relax
  \def\url#1{\burl{#1}}\fi
\providecommand{\bibinfo}[2]{#2}
\providecommand{\eprint}[2][]{\url{#2}}
\providecommand{\doi}[1]{\url{https://doi.org/#1}}

\bibitem{shi_recent_2017}
\bibinfo{author}{Shi, R.}, \bibinfo{author}{Waterhouse, G.~I.} \&
  \bibinfo{author}{Zhang, T.}
\newblock \bibinfo{title}{Recent {Progress} in {Photocatalytic} {CO2}
  {Reduction} {Over} {Perovskite} {Oxides}}.
\newblock \emph{\bibinfo{journal}{Solar RRL}}
  \textbf{\bibinfo{volume}{1}}~(11), \bibinfo{pages}{1700126}
  (\bibinfo{year}{2017}).
\newblock \doi{10.1002/solr.201700126}.

\bibitem{sumaria_atomic-scale_2023}
\bibinfo{author}{Sumaria, V.}, \bibinfo{author}{Nguyen, L.},
  \bibinfo{author}{Tao, F.~F.} \& \bibinfo{author}{Sautet, P.}
\newblock \bibinfo{title}{Atomic-{Scale} {Mechanism} of {Platinum} {Catalyst}
  {Restructuring} under a {Pressure} of {Reactant} {Gas}}.
\newblock \emph{\bibinfo{journal}{Journal of the American Chemical Society}}
  \textbf{\bibinfo{volume}{145}}~(1), \bibinfo{pages}{392--401}
  (\bibinfo{year}{2023}).
\newblock \doi{10.1021/jacs.2c10179}.

\bibitem{fabbri_dynamic_2017}
\bibinfo{author}{Fabbri, E.} \emph{et~al.}
\newblock \bibinfo{title}{Dynamic surface self-reconstruction is the key of
  highly active perovskite nano-electrocatalysts for water splitting}.
\newblock \emph{\bibinfo{journal}{Nature Materials}}
  \textbf{\bibinfo{volume}{16}}~(9), \bibinfo{pages}{925--931}
  (\bibinfo{year}{2017}).
\newblock \doi{10.1038/nmat4938}.

\bibitem{zhang_hydrogen-induced_2022}
\bibinfo{author}{Zhang, Z.}, \bibinfo{author}{Wei, Z.},
  \bibinfo{author}{Sautet, P.} \& \bibinfo{author}{Alexandrova, A.~N.}
\newblock \bibinfo{title}{Hydrogen-{Induced} {Restructuring} of a {Cu}(100)
  {Electrode} in {Electroreduction} {Conditions}}.
\newblock \emph{\bibinfo{journal}{Journal of the American Chemical Society}}
  \textbf{\bibinfo{volume}{144}}~(42), \bibinfo{pages}{19284--19293}
  (\bibinfo{year}{2022}).
\newblock \doi{10.1021/jacs.2c06188}.

\bibitem{sha_understanding_2023}
\bibinfo{author}{Sha, Z.}, \bibinfo{author}{Shen, Z.}, \bibinfo{author}{Calì,
  E.}, \bibinfo{author}{Kilner, J.~A.} \& \bibinfo{author}{Skinner, S.~J.}
\newblock \bibinfo{title}{Understanding surface chemical processes in
  perovskite oxide electrodes}.
\newblock \emph{\bibinfo{journal}{Journal of Materials Chemistry A}}
  (\bibinfo{year}{2023}).
\newblock \doi{10.1039/D3TA00070B}.

\bibitem{jung_understanding_2014}
\bibinfo{author}{Jung, S.-K.} \emph{et~al.}
\newblock \bibinfo{title}{Understanding the {Degradation} {Mechanisms} of
  {LiNi0}.{5Co0}.{2Mn0}.{3O2} {Cathode} {Material} in {Lithium} {Ion}
  {Batteries}}.
\newblock \emph{\bibinfo{journal}{Advanced Energy Materials}}
  \textbf{\bibinfo{volume}{4}}~(1), \bibinfo{pages}{1300787}
  (\bibinfo{year}{2014}).
\newblock \doi{10.1002/aenm.201300787}.

\bibitem{han_coating_2017}
\bibinfo{author}{Han, B.} \emph{et~al.}
\newblock \bibinfo{title}{From {Coating} to {Dopant}: {How} the {Transition}
  {Metal} {Composition} {Affects} {Alumina} {Coatings} on {Ni}-{Rich}
  {Cathodes}}.
\newblock \emph{\bibinfo{journal}{ACS Applied Materials \& Interfaces}}
  \textbf{\bibinfo{volume}{9}}~(47), \bibinfo{pages}{41291--41302}
  (\bibinfo{year}{2017}).
\newblock \doi{10.1021/acsami.7b13597}.

\bibitem{xu_bulk_2021}
\bibinfo{author}{Xu, C.} \emph{et~al.}
\newblock \bibinfo{title}{Bulk fatigue induced by surface reconstruction in
  layered {Ni}-rich cathodes for {Li}-ion batteries}.
\newblock \emph{\bibinfo{journal}{Nature Materials}}
  \textbf{\bibinfo{volume}{20}}~(1), \bibinfo{pages}{84--92}
  (\bibinfo{year}{2021}).
\newblock \doi{10.1038/s41563-020-0767-8}.

\bibitem{hirata_electronic_1994}
\bibinfo{author}{Hirata, A.}, \bibinfo{author}{Saiki, K.},
  \bibinfo{author}{Koma, A.} \& \bibinfo{author}{Ando, A.}
\newblock \bibinfo{title}{Electronic structure of a {SrO}-terminated
  {SrTiO3}(100) surface}.
\newblock \emph{\bibinfo{journal}{Surface Science}}
  \textbf{\bibinfo{volume}{319}}~(3), \bibinfo{pages}{267--271}
  (\bibinfo{year}{1994}).
\newblock \doi{10.1016/0039-6028(94)90593-2}.

\bibitem{castell_scanning_2002}
\bibinfo{author}{Castell, M.~R.}
\newblock \bibinfo{title}{Scanning tunneling microscopy of reconstructions on
  the {SrTiO3}(001) surface}.
\newblock \emph{\bibinfo{journal}{Surface Science}}
  \textbf{\bibinfo{volume}{505}}, \bibinfo{pages}{1--13}
  (\bibinfo{year}{2002}).
\newblock \doi{10.1016/S0039-6028(02)01393-6}.

\bibitem{erdman_structure_2002}
\bibinfo{author}{Erdman, N.} \emph{et~al.}
\newblock \bibinfo{title}{The structure and chemistry of the {TiO2}-rich
  surface of {SrTiO3}(001)}.
\newblock \emph{\bibinfo{journal}{Nature}}
  \textbf{\bibinfo{volume}{419}}~(6902), \bibinfo{pages}{55--58}
  (\bibinfo{year}{2002}).
\newblock \doi{10.1038/nature01010}.

\bibitem{heifets_electronic_2007}
\bibinfo{author}{Heifets, E.}, \bibinfo{author}{Piskunov, S.},
  \bibinfo{author}{Kotomin, E.~A.}, \bibinfo{author}{Zhukovskii, Y.~F.} \&
  \bibinfo{author}{Ellis, D.~E.}
\newblock \bibinfo{title}{Electronic structure and thermodynamic stability of
  double-layered {SrTiO3}(001) surfaces: \textit{{Ab} initio} simulations}.
\newblock \emph{\bibinfo{journal}{Physical Review B}}
  \textbf{\bibinfo{volume}{75}}~(11), \bibinfo{pages}{115417}
  (\bibinfo{year}{2007}).
\newblock \doi{10.1103/PhysRevB.75.115417}.

\bibitem{li_data-driven_2023}
\bibinfo{author}{Li, H.}, \bibinfo{author}{Jiao, Y.}, \bibinfo{author}{Davey,
  K.} \& \bibinfo{author}{Qiao, S.-Z.}
\newblock \bibinfo{title}{Data-{Driven} {Machine} {Learning} for
  {Understanding} {Surface} {Structures} of {Heterogeneous} {Catalysts}}.
\newblock \emph{\bibinfo{journal}{Angewandte Chemie}}
  \textbf{\bibinfo{volume}{135}}~(9), \bibinfo{pages}{e202216383}
  (\bibinfo{year}{2023}).
\newblock \doi{10.1002/ange.202216383}.

\bibitem{merte_structure_2022}
\bibinfo{author}{Merte, L.~R.} \emph{et~al.}
\newblock \bibinfo{title}{Structure of an {Ultrathin} {Oxide} on {Pt3Sn}(111)
  {Solved} by {Machine} {Learning} {Enhanced} {Global} {Optimization}}.
\newblock \emph{\bibinfo{journal}{Angewandte Chemie International Edition}}
  \textbf{\bibinfo{volume}{61}}~(25), \bibinfo{pages}{e202204244}
  (\bibinfo{year}{2022}).
\newblock \doi{10.1002/anie.202204244}.

\bibitem{foiles_embedded-atom-method_1986}
\bibinfo{author}{Foiles, S.~M.}, \bibinfo{author}{Baskes, M.~I.} \&
  \bibinfo{author}{Daw, M.~S.}
\newblock \bibinfo{title}{Embedded-atom-method functions for the fcc metals
  {Cu}, {Ag}, {Au}, {Ni}, {Pd}, {Pt}, and their alloys}.
\newblock \emph{\bibinfo{journal}{Physical Review B}}
  \textbf{\bibinfo{volume}{33}}~(12), \bibinfo{pages}{7983--7991}
  (\bibinfo{year}{1986}).
\newblock \doi{10.1103/PhysRevB.33.7983}.

\bibitem{nord_modelling_2003}
\bibinfo{author}{Nord, J.}, \bibinfo{author}{Albe, K.},
  \bibinfo{author}{Erhart, P.} \& \bibinfo{author}{Nordlund, K.}
\newblock \bibinfo{title}{Modelling of compound semiconductors: analytical
  bond-order potential for gallium, nitrogen and gallium nitride}.
\newblock \emph{\bibinfo{journal}{Journal of Physics: Condensed Matter}}
  \textbf{\bibinfo{volume}{15}}~(32), \bibinfo{pages}{5649}
  (\bibinfo{year}{2003}).
\newblock \doi{10.1088/0953-8984/15/32/324}.

\bibitem{kolpak_evolution_2008}
\bibinfo{author}{Kolpak, A.~M.}, \bibinfo{author}{Li, D.},
  \bibinfo{author}{Shao, R.}, \bibinfo{author}{Rappe, A.~M.} \&
  \bibinfo{author}{Bonnell, D.~A.}
\newblock \bibinfo{title}{Evolution of the {Structure} and {Thermodynamic}
  {Stability} of the {BaTiO3}(001) {Surface}}.
\newblock \emph{\bibinfo{journal}{Physical Review Letters}}
  \textbf{\bibinfo{volume}{101}}~(3), \bibinfo{pages}{036102}
  (\bibinfo{year}{2008}).
\newblock \doi{10.1103/PhysRevLett.101.036102}.

\bibitem{wexler_automatic_2019}
\bibinfo{author}{Wexler, R.~B.}, \bibinfo{author}{Qiu, T.} \&
  \bibinfo{author}{Rappe, A.~M.}
\newblock \bibinfo{title}{Automatic {Prediction} of {Surface} {Phase}
  {Diagrams} {Using} {Ab} {Initio} {Grand} {Canonical} {Monte} {Carlo}}.
\newblock \emph{\bibinfo{journal}{The Journal of Physical Chemistry C}}
  \textbf{\bibinfo{volume}{123}}~(4), \bibinfo{pages}{2321--2328}
  (\bibinfo{year}{2019}).
\newblock \doi{10.1021/acs.jpcc.8b11093}.

\bibitem{zhou_unexpected_2014}
\bibinfo{author}{Zhou, X.-F.}, \bibinfo{author}{Oganov, A.~R.},
  \bibinfo{author}{Shao, X.}, \bibinfo{author}{Zhu, Q.} \&
  \bibinfo{author}{Wang, H.-T.}
\newblock \bibinfo{title}{Unexpected {Reconstruction} of the
  {\textalpha}-{Boron} (111) {Surface}}.
\newblock \emph{\bibinfo{journal}{Physical Review Letters}}
  \textbf{\bibinfo{volume}{113}}~(17), \bibinfo{pages}{176101}
  (\bibinfo{year}{2014}).
\newblock \doi{10.1103/PhysRevLett.113.176101}.

\bibitem{timmermann_iro2_2020}
\bibinfo{author}{Timmermann, J.} \emph{et~al.}
\newblock \bibinfo{title}{{IrO2} {Surface} {Complexions} {Identified} through
  {Machine} {Learning} and {Surface} {Investigations}}.
\newblock \emph{\bibinfo{journal}{Physical Review Letters}}
  \textbf{\bibinfo{volume}{125}}~(20), \bibinfo{pages}{206101}
  (\bibinfo{year}{2020}).
\newblock \doi{10.1103/PhysRevLett.125.206101}.

\bibitem{wales_global_1997}
\bibinfo{author}{Wales, D.~J.} \& \bibinfo{author}{Doye, J. P.~K.}
\newblock \bibinfo{title}{Global {Optimization} by {Basin}-{Hopping} and the
  {Lowest} {Energy} {Structures} of {Lennard}-{Jones} {Clusters} {Containing}
  up to 110 {Atoms}}.
\newblock \emph{\bibinfo{journal}{The Journal of Physical Chemistry A}}
  \textbf{\bibinfo{volume}{101}}~(28), \bibinfo{pages}{5111--5116}
  (\bibinfo{year}{1997}).
\newblock \doi{10.1021/jp970984n}.

\bibitem{panosetti_global_2015}
\bibinfo{author}{Panosetti, C.}, \bibinfo{author}{Krautgasser, K.},
  \bibinfo{author}{Palagin, D.}, \bibinfo{author}{Reuter, K.} \&
  \bibinfo{author}{Maurer, R.~J.}
\newblock \bibinfo{title}{Global {Materials} {Structure} {Search} with
  {Chemically} {Motivated} {Coordinates}}.
\newblock \emph{\bibinfo{journal}{Nano Letters}}
  \textbf{\bibinfo{volume}{15}}~(12), \bibinfo{pages}{8044--8048}
  (\bibinfo{year}{2015}).
\newblock \doi{10.1021/acs.nanolett.5b03388}.

\bibitem{obersteiner_structure_2017}
\bibinfo{author}{Obersteiner, V.}, \bibinfo{author}{Scherbela, M.},
  \bibinfo{author}{Hörmann, L.}, \bibinfo{author}{Wegner, D.} \&
  \bibinfo{author}{Hofmann, O.~T.}
\newblock \bibinfo{title}{Structure {Prediction} for {Surface}-{Induced}
  {Phases} of {Organic} {Monolayers}: {Overcoming} the {Combinatorial}
  {Bottleneck}}.
\newblock \emph{\bibinfo{journal}{Nano Letters}}
  \textbf{\bibinfo{volume}{17}}~(7), \bibinfo{pages}{4453--4460}
  (\bibinfo{year}{2017}).
\newblock \doi{10.1021/acs.nanolett.7b01637}.

\bibitem{egger_charge_2020}
\bibinfo{author}{Egger, A.~T.} \emph{et~al.}
\newblock \bibinfo{title}{Charge {Transfer} into {Organic} {Thin} {Films}: {A}
  {Deeper} {Insight} through {Machine}-{Learning}-{Assisted} {Structure}
  {Search}}.
\newblock \emph{\bibinfo{journal}{Advanced Science}}
  \textbf{\bibinfo{volume}{7}}~(15), \bibinfo{pages}{2000992}
  (\bibinfo{year}{2020}).
\newblock \doi{10.1002/advs.202000992}.

\bibitem{bauer_systematic_2022}
\bibinfo{author}{Bauer, M.~N.}, \bibinfo{author}{Probert, M. I.~J.} \&
  \bibinfo{author}{Panosetti, C.}
\newblock \bibinfo{title}{Systematic {Comparison} of {Genetic} {Algorithm} and
  {Basin} {Hopping} {Approaches} to the {Global} {Optimization} of {Si}(111)
  {Surface} {Reconstructions}}.
\newblock \emph{\bibinfo{journal}{The Journal of Physical Chemistry A}}
  \textbf{\bibinfo{volume}{126}}~(19), \bibinfo{pages}{3043--3056}
  (\bibinfo{year}{2022}).
\newblock \doi{10.1021/acs.jpca.2c00647}.

\bibitem{wang_new_2014}
\bibinfo{author}{Wang, Q.}, \bibinfo{author}{Oganov, A.~R.},
  \bibinfo{author}{Zhu, Q.} \& \bibinfo{author}{Zhou, X.-F.}
\newblock \bibinfo{title}{New {Reconstructions} of the (110) {Surface} of
  {Rutile} {TiO2} {Predicted} by an {Evolutionary} {Method}}.
\newblock \emph{\bibinfo{journal}{Physical Review Letters}}
  \textbf{\bibinfo{volume}{113}}~(26), \bibinfo{pages}{266101}
  (\bibinfo{year}{2014}).
\newblock \doi{10.1103/PhysRevLett.113.266101}.

\bibitem{schusteritsch_predicting_2014}
\bibinfo{author}{Schusteritsch, G.} \& \bibinfo{author}{Pickard, C.~J.}
\newblock \bibinfo{title}{Predicting interface structures: {From} {SrTiO3} to
  graphene}.
\newblock \emph{\bibinfo{journal}{Physical Review B}}
  \textbf{\bibinfo{volume}{90}}~(3), \bibinfo{pages}{035424}
  (\bibinfo{year}{2014}).
\newblock \doi{10.1103/PhysRevB.90.035424}.

\bibitem{meldgaard_structure_2020}
\bibinfo{author}{Meldgaard, S.~A.}, \bibinfo{author}{Mortensen, H.~L.},
  \bibinfo{author}{Jørgensen, M.~S.} \& \bibinfo{author}{Hammer, B.}
\newblock \bibinfo{title}{Structure prediction of surface reconstructions by
  deep reinforcement learning}.
\newblock \emph{\bibinfo{journal}{Journal of Physics: Condensed Matter}}
  \textbf{\bibinfo{volume}{32}}~(40), \bibinfo{pages}{404005}
  (\bibinfo{year}{2020}).
\newblock \doi{10.1088/1361-648X/ab94f2}.

\bibitem{hess_polar_2020}
\bibinfo{author}{Hess, F.} \& \bibinfo{author}{Yildiz, B.}
\newblock \bibinfo{title}{Polar or not polar? {The} interplay between
  reconstruction, {Sr} enrichment, and reduction at the
  {La0}.{75Sr0}.{25MnO3}(001) surface}.
\newblock \emph{\bibinfo{journal}{Physical Review Materials}}
  \textbf{\bibinfo{volume}{4}}~(1), \bibinfo{pages}{015801}
  (\bibinfo{year}{2020}).
\newblock \doi{10.1103/PhysRevMaterials.4.015801}.

\bibitem{unke_machine_2021}
\bibinfo{author}{Unke, O.~T.} \emph{et~al.}
\newblock \bibinfo{title}{Machine {Learning} {Force} {Fields}}.
\newblock \emph{\bibinfo{journal}{Chemical Reviews}}
  \textbf{\bibinfo{volume}{121}}~(16), \bibinfo{pages}{10142--10186}
  (\bibinfo{year}{2021}).
\newblock \doi{10.1021/acs.chemrev.0c01111}.

\bibitem{axelrod_learning_2022}
\bibinfo{author}{Axelrod, S.} \emph{et~al.}
\newblock \bibinfo{title}{Learning {Matter}: {Materials} {Design} with
  {Machine} {Learning} and {Atomistic} {Simulations}}.
\newblock \emph{\bibinfo{journal}{Accounts of Materials Research}}
  \textbf{\bibinfo{volume}{3}}~(3), \bibinfo{pages}{343--357}
  (\bibinfo{year}{2022}).
\newblock \doi{10.1021/accountsmr.1c00238}.

\bibitem{bisbo_efficient_2020}
\bibinfo{author}{Bisbo, M.~K.} \& \bibinfo{author}{Hammer, B.}
\newblock \bibinfo{title}{Efficient {Global} {Structure} {Optimization} with a
  {Machine}-{Learned} {Surrogate} {Model}}.
\newblock \emph{\bibinfo{journal}{Physical Review Letters}}
  \textbf{\bibinfo{volume}{124}}~(8), \bibinfo{pages}{086102}
  (\bibinfo{year}{2020}).
\newblock \doi{10.1103/PhysRevLett.124.086102}.

\bibitem{bisbo_global_2022}
\bibinfo{author}{Bisbo, M.~K.} \& \bibinfo{author}{Hammer, B.}
\newblock \bibinfo{title}{Global optimization of atomic structure enhanced by
  machine learning}.
\newblock \emph{\bibinfo{journal}{Physical Review B}}
  \textbf{\bibinfo{volume}{105}}~(24), \bibinfo{pages}{245404}
  (\bibinfo{year}{2022}).
\newblock \doi{10.1103/PhysRevB.105.245404}.

\bibitem{timmermann_data-efficient_2021}
\bibinfo{author}{Timmermann, J.} \emph{et~al.}
\newblock \bibinfo{title}{Data-efficient iterative training of {Gaussian}
  approximation potentials: {Application} to surface structure determination of
  rutile {IrO2} and {RuO2}}.
\newblock \emph{\bibinfo{journal}{The Journal of Chemical Physics}}
  \textbf{\bibinfo{volume}{155}}~(24), \bibinfo{pages}{244107}
  (\bibinfo{year}{2021}).
\newblock \doi{10.1063/5.0071249}.

\bibitem{ronne_atomistic_2022}
\bibinfo{author}{Rønne, N.} \emph{et~al.}
\newblock \bibinfo{title}{Atomistic structure search using local surrogate
  model}.
\newblock \emph{\bibinfo{journal}{The Journal of Chemical Physics}}
  \textbf{\bibinfo{volume}{157}}~(17), \bibinfo{pages}{174115}
  (\bibinfo{year}{2022}).
\newblock \doi{10.1063/5.0121748}.

\bibitem{han_prediction_2023}
\bibinfo{author}{Han, Y.} \emph{et~al.}
\newblock \bibinfo{title}{Prediction of surface reconstructions using {MAGUS}}.
\newblock \emph{\bibinfo{journal}{The Journal of Chemical Physics}}
  \textbf{\bibinfo{volume}{158}}~(17), \bibinfo{pages}{174109}
  (\bibinfo{year}{2023}).
\newblock \doi{10.1063/5.0142281}.

\bibitem{xu_atomistic_2022}
\bibinfo{author}{Xu, J.}, \bibinfo{author}{Xie, W.}, \bibinfo{author}{Han, Y.}
  \& \bibinfo{author}{Hu, P.}
\newblock \bibinfo{title}{Atomistic {Insights} into the {Oxidation} of {Flat}
  and {Stepped} {Platinum} {Surfaces} {Using} {Large}-{Scale} {Machine}
  {Learning} {Potential}-{Based} {Grand}-{Canonical} {Monte} {Carlo}}.
\newblock \emph{\bibinfo{journal}{ACS Catalysis}}
  \textbf{\bibinfo{volume}{12}}~(24), \bibinfo{pages}{14812--14824}
  (\bibinfo{year}{2022}).
\newblock \doi{10.1021/acscatal.2c03976}.

\bibitem{bernardin_semi-grand_2007}
\bibinfo{author}{Bernardin, F.~E.} \& \bibinfo{author}{Rutledge, G.~C.}
\newblock \bibinfo{title}{Semi-{Grand} {Canonical} {Monte} {Carlo} ({SGMC})
  {Simulations} to {Interpret} {Experimental} {Data} on {Processed} {Polymer}
  {Melts} and {Glasses}}.
\newblock \emph{\bibinfo{journal}{Macromolecules}}
  \textbf{\bibinfo{volume}{40}}~(13), \bibinfo{pages}{4691--4702}
  (\bibinfo{year}{2007}).
\newblock \doi{10.1021/ma062935r}.

\bibitem{damewood_sampling_2022}
\bibinfo{author}{Damewood, J.}, \bibinfo{author}{Schwalbe-Koda, D.} \&
  \bibinfo{author}{Gómez-Bombarelli, R.}
\newblock \bibinfo{title}{Sampling lattices in semi-grand canonical ensemble
  with autoregressive machine learning}.
\newblock \emph{\bibinfo{journal}{npj Computational Materials}}
  \textbf{\bibinfo{volume}{8}}~(1), \bibinfo{pages}{1--10}
  (\bibinfo{year}{2022}).
\newblock \doi{10.1038/s41524-022-00736-4}.

\bibitem{carrete_deep_2023}
\bibinfo{author}{Carrete, J.}, \bibinfo{author}{Montes-Campos, H.},
  \bibinfo{author}{Wanzenböck, R.}, \bibinfo{author}{Heid, E.} \&
  \bibinfo{author}{Madsen, G. K.~H.}
\newblock \bibinfo{title}{Deep ensembles vs committees for uncertainty
  estimation in neural-network force fields: {Comparison} and application to
  active learning}.
\newblock \emph{\bibinfo{journal}{The Journal of Chemical Physics}}
  \textbf{\bibinfo{volume}{158}}~(20), \bibinfo{pages}{204801}
  (\bibinfo{year}{2023}).
\newblock \doi{10.1063/5.0146905}.

\bibitem{tan_single-model_2023}
\bibinfo{author}{Tan, A.~R.}, \bibinfo{author}{Urata, S.},
  \bibinfo{author}{Goldman, S.}, \bibinfo{author}{Dietschreit, J. C.~B.} \&
  \bibinfo{author}{Gómez-Bombarelli, R.}
\newblock \bibinfo{title}{Single-model uncertainty quantification in neural
  network potentials does not consistently outperform model ensembles}
  (\bibinfo{year}{2023}).
\newblock \bibinfo{note}{ArXiv:2305.01754 [physics]}.

\bibitem{schwalbe-koda_differentiable_2021}
\bibinfo{author}{Schwalbe-Koda, D.}, \bibinfo{author}{Tan, A.~R.} \&
  \bibinfo{author}{Gómez-Bombarelli, R.}
\newblock \bibinfo{title}{Differentiable sampling of molecular geometries with
  uncertainty-based adversarial attacks}.
\newblock \emph{\bibinfo{journal}{Nature Communications}}
  \textbf{\bibinfo{volume}{12}}~(1), \bibinfo{pages}{5104}
  (\bibinfo{year}{2021}).
\newblock \doi{10.1038/s41467-021-25342-8}.

\bibitem{fu_forces_2022}
\bibinfo{author}{Fu, X.} \emph{et~al.}
\newblock \bibinfo{title}{Forces are not {Enough}: {Benchmark} and {Critical}
  {Evaluation} for {Machine} {Learning} {Force} {Fields} with {Molecular}
  {Simulations}} (\bibinfo{year}{2022}).
\newblock \bibinfo{note}{ArXiv:2210.07237 [physics]}.

\bibitem{damewood_representations_2023}
\bibinfo{author}{Damewood, J.} \emph{et~al.}
\newblock \bibinfo{title}{Representations of materials for machine learning}.
\newblock \emph{\bibinfo{journal}{Annual Review of Materials Research}}
  \textbf{\bibinfo{volume}{53}}~(1), \bibinfo{pages}{399--426}
  (\bibinfo{year}{2023}).
\newblock \doi{10.1146/annurev-matsci-080921-085947}.

\bibitem{stephenson_modified_1996}
\bibinfo{author}{Stephenson, P. C.~L.}, \bibinfo{author}{Radny, M.~W.} \&
  \bibinfo{author}{Smith, P.~V.}
\newblock \bibinfo{title}{A modified {Stillinger}-{Weber} potential for
  modelling silicon surfaces}.
\newblock \emph{\bibinfo{journal}{Surface Science}}
  \textbf{\bibinfo{volume}{366}}~(1), \bibinfo{pages}{177--184}
  (\bibinfo{year}{1996}).
\newblock \doi{10.1016/0039-6028(96)00801-1}.

\bibitem{northrup_structure_2000}
\bibinfo{author}{Northrup, J.~E.}, \bibinfo{author}{Neugebauer, J.},
  \bibinfo{author}{Feenstra, R.~M.} \& \bibinfo{author}{Smith, A.~R.}
\newblock \bibinfo{title}{Structure of {GaN}(0001): {The} laterally contracted
  {Ga} bilayer model}.
\newblock \emph{\bibinfo{journal}{Physical Review B}}
  \textbf{\bibinfo{volume}{61}}~(15), \bibinfo{pages}{9932--9935}
  (\bibinfo{year}{2000}).
\newblock \doi{10.1103/PhysRevB.61.9932}.

\bibitem{stich_ab_1992}
\bibinfo{author}{Štich, I.}, \bibinfo{author}{Payne, M.~C.},
  \bibinfo{author}{King-Smith, R.~D.}, \bibinfo{author}{Lin, J.-S.} \&
  \bibinfo{author}{Clarke, L.~J.}
\newblock \bibinfo{title}{Ab initio total-energy calculations for extremely
  large systems: {Application} to the {Takayanagi} reconstruction of
  {Si}(111)}.
\newblock \emph{\bibinfo{journal}{Physical Review Letters}}
  \textbf{\bibinfo{volume}{68}}~(9), \bibinfo{pages}{1351--1354}
  (\bibinfo{year}{1992}).
\newblock \doi{10.1103/PhysRevLett.68.1351}.

\bibitem{smeu_electronic_2012}
\bibinfo{author}{Smeu, M.}, \bibinfo{author}{Guo, H.}, \bibinfo{author}{Ji, W.}
  \& \bibinfo{author}{Wolkow, R.~A.}
\newblock \bibinfo{title}{Electronic properties of {Si}(111)-7x7 and related
  reconstructions: {Density} functional theory calculations}.
\newblock \emph{\bibinfo{journal}{Physical Review B}}
  \textbf{\bibinfo{volume}{85}}~(19), \bibinfo{pages}{195315}
  (\bibinfo{year}{2012}).
\newblock \doi{10.1103/PhysRevB.85.195315}.

\bibitem{herger_surface_2007}
\bibinfo{author}{Herger, R.} \emph{et~al.}
\newblock \bibinfo{title}{Surface of {Strontium} {Titanate}}.
\newblock \emph{\bibinfo{journal}{Physical Review Letters}}
  \textbf{\bibinfo{volume}{98}}~(7), \bibinfo{pages}{076102}
  (\bibinfo{year}{2007}).
\newblock \doi{10.1103/PhysRevLett.98.076102}.

\bibitem{hong_anomalous_2023}
\bibinfo{author}{Hong, C.} \emph{et~al.}
\newblock \bibinfo{title}{Anomalous intense coherent secondary photoemission
  from a perovskite oxide}.
\newblock \emph{\bibinfo{journal}{Nature}} \bibinfo{pages}{1--3}
  (\bibinfo{year}{2023}).
\newblock \doi{10.1038/s41586-023-05900-4}, \bibinfo{note}{publisher: Nature
  Publishing Group}.

\bibitem{szot_surfaces_1999}
\bibinfo{author}{Szot, K.} \& \bibinfo{author}{Speier, W.}
\newblock \bibinfo{title}{Surfaces of reduced and oxidized {SrTiO3} from atomic
  force microscopy}.
\newblock \emph{\bibinfo{journal}{Physical Review B}}
  \textbf{\bibinfo{volume}{60}}~(8), \bibinfo{pages}{5909--5926}
  (\bibinfo{year}{1999}).
\newblock \doi{10.1103/PhysRevB.60.5909}.

\bibitem{kubo_surface_2003}
\bibinfo{author}{Kubo, T.} \& \bibinfo{author}{Nozoye, H.}
\newblock \bibinfo{title}{Surface structure of {SrTiO3}(100)}.
\newblock \emph{\bibinfo{journal}{Surface Science}}
  \textbf{\bibinfo{volume}{542}}~(3), \bibinfo{pages}{177--191}
  (\bibinfo{year}{2003}).
\newblock \doi{10.1016/S0039-6028(03)00998-1}.

\bibitem{winter_simulations_2023}
\bibinfo{author}{Winter, G.} \& \bibinfo{author}{Gómez-Bombarelli, R.}
\newblock \bibinfo{title}{Simulations with machine learning potentials identify
  the ion conduction mechanism mediating non-{Arrhenius} behavior in {LGPS}}.
\newblock \emph{\bibinfo{journal}{Journal of Physics: Energy}}
  \textbf{\bibinfo{volume}{5}}~(2), \bibinfo{pages}{024004}
  (\bibinfo{year}{2023}).
\newblock \doi{10.1088/2515-7655/acbbef}.

\bibitem{millan_effect_2023}
\bibinfo{author}{Millan, R.}, \bibinfo{author}{Bello-Jurado, E.},
  \bibinfo{author}{Moliner, M.}, \bibinfo{author}{Boronat, M.} \&
  \bibinfo{author}{Gomez-Bombarelli, R.}
\newblock \bibinfo{title}{Effect of {Framework} {Composition} and {NH3} on the
  {Diffusion} of {Cu}+ in {Cu}-{CHA} {Catalysts} {Predicted} by
  {Machine}-{Learning} {Accelerated} {Molecular} {Dynamics}}.
\newblock \emph{\bibinfo{journal}{ACS Central Science}}
  (\bibinfo{year}{2023}).
\newblock \doi{10.1021/acscentsci.3c00870}.

\bibitem{thompson_lammps_2022}
\bibinfo{author}{Thompson, A.~P.} \emph{et~al.}
\newblock \bibinfo{title}{{LAMMPS} - a flexible simulation tool for
  particle-based materials modeling at the atomic, meso, and continuum scales}.
\newblock \emph{\bibinfo{journal}{Computer Physics Communications}}
  \textbf{\bibinfo{volume}{271}}, \bibinfo{pages}{108171}
  (\bibinfo{year}{2022}).
\newblock \doi{10.1016/j.cpc.2021.108171}.

\bibitem{larsen_atomic_2017}
\bibinfo{author}{Larsen, A.~H.} \emph{et~al.}
\newblock \bibinfo{title}{The atomic simulation environment—a {Python}
  library for working with atoms}.
\newblock \emph{\bibinfo{journal}{Journal of Physics: Condensed Matter}}
  \textbf{\bibinfo{volume}{29}}~(27), \bibinfo{pages}{273002}
  (\bibinfo{year}{2017}).
\newblock \doi{10.1088/1361-648X/aa680e}.

\bibitem{boes_graph_2019}
\bibinfo{author}{Boes, J.~R.}, \bibinfo{author}{Mamun, O.},
  \bibinfo{author}{Winther, K.} \& \bibinfo{author}{Bligaard, T.}
\newblock \bibinfo{title}{Graph {Theory} {Approach} to {High}-{Throughput}
  {Surface} {Adsorption} {Structure} {Generation}}.
\newblock \emph{\bibinfo{journal}{The Journal of Physical Chemistry A}}
  \textbf{\bibinfo{volume}{123}}~(11), \bibinfo{pages}{2281--2285}
  (\bibinfo{year}{2019}).
\newblock \doi{10.1021/acs.jpca.9b00311}.

\bibitem{ong_python_2013}
\bibinfo{author}{Ong, S.~P.} \emph{et~al.}
\newblock \bibinfo{title}{Python {Materials} {Genomics} (pymatgen): {A} robust,
  open-source python library for materials analysis}.
\newblock \emph{\bibinfo{journal}{Computational Materials Science}}
  \textbf{\bibinfo{volume}{68}}, \bibinfo{pages}{314--319}
  (\bibinfo{year}{2013}).
\newblock \doi{10.1016/j.commatsci.2012.10.028}.

\bibitem{momma_vesta_2011}
\bibinfo{author}{Momma, K.} \& \bibinfo{author}{Izumi, F.}
\newblock \bibinfo{title}{{VESTA} 3 for three-dimensional visualization of
  crystal, volumetric and morphology data}.
\newblock \emph{\bibinfo{journal}{Journal of Applied Crystallography}}
  \textbf{\bibinfo{volume}{44}}~(6), \bibinfo{pages}{1272--1276}
  (\bibinfo{year}{2011}).
\newblock \doi{10.1107/S0021889811038970}.

\bibitem{jain_commentary_2013}
\bibinfo{author}{Jain, A.} \emph{et~al.}
\newblock \bibinfo{title}{Commentary: {The} {Materials} {Project}: {A}
  materials genome approach to accelerating materials innovation}.
\newblock \emph{\bibinfo{journal}{APL Materials}}
  \textbf{\bibinfo{volume}{1}}~(1), \bibinfo{pages}{011002}
  (\bibinfo{year}{2013}).
\newblock \doi{10.1063/1.4812323}.

\bibitem{schutt_equivariant_2021}
\bibinfo{author}{Schütt, K.}, \bibinfo{author}{Unke, O.} \&
  \bibinfo{author}{Gastegger, M.}
\newblock \bibinfo{editor}{Meila, M.} \& \bibinfo{editor}{Zhang, T.} (eds)
  \emph{\bibinfo{title}{Equivariant message passing for the prediction of
  tensorial properties and molecular spectra}}.
\newblock (eds \bibinfo{editor}{Meila, M.} \& \bibinfo{editor}{Zhang, T.})
  \emph{\bibinfo{booktitle}{Proceedings of the 38th international conference on
  machine learning}}, Vol. \bibinfo{volume}{139} of
  \emph{\bibinfo{series}{Proceedings of {Machine} {Learning} {Research}}},
  \bibinfo{pages}{9377--9388} (\bibinfo{publisher}{PMLR},
  \bibinfo{address}{Virtual}, \bibinfo{year}{2021}).

\bibitem{martinez-cantin_practical_2018}
\bibinfo{author}{Martinez-Cantin, R.}, \bibinfo{author}{Tee, K.} \&
  \bibinfo{author}{McCourt, M.}
\newblock \bibinfo{editor}{Storkey, A.} \& \bibinfo{editor}{Perez-Cruz, F.}
  (eds) \emph{\bibinfo{title}{Practical {Bayesian} optimization in the presence
  of outliers}}.
\newblock (eds \bibinfo{editor}{Storkey, A.} \& \bibinfo{editor}{Perez-Cruz,
  F.}) \emph{\bibinfo{booktitle}{Proceedings of the {Twenty}-{First}
  {International} {Conference} on {Artificial} {Intelligence} and
  {Statistics}}}, Vol.~\bibinfo{volume}{84} of
  \emph{\bibinfo{series}{Proceedings of {Machine} {Learning} {Research}}},
  \bibinfo{pages}{1722--1731} (\bibinfo{publisher}{PMLR},
  \bibinfo{address}{Playa Blanca, Lanzarote, Canary Islands},
  \bibinfo{year}{2018}).

\bibitem{ramachandran_searching_2017}
\bibinfo{author}{Ramachandran, P.}, \bibinfo{author}{Zoph, B.} \&
  \bibinfo{author}{Le, Q.~V.}
\newblock \bibinfo{title}{Searching for {Activation} {Functions}}
  (\bibinfo{year}{2017}).
\newblock \bibinfo{note}{ArXiv:1710.05941 [cs] version: 2}.

\bibitem{kingma_adam_2015}
\bibinfo{author}{Kingma, D.~P.} \& \bibinfo{author}{Ba, J.}
\newblock \bibinfo{editor}{Bengio, Y.} \& \bibinfo{editor}{LeCun, Y.} (eds)
  \emph{\bibinfo{title}{Adam: {A} {Method} for {Stochastic} {Optimization}}}.
\newblock (eds \bibinfo{editor}{Bengio, Y.} \& \bibinfo{editor}{LeCun, Y.})
  \emph{\bibinfo{booktitle}{3rd {International} {Conference} on {Learning}
  {Representations}, {ICLR} 2015, {San} {Diego}, {CA}, {USA}, {May} 7-9, 2015,
  {Conference} {Track} {Proceedings}}} (\bibinfo{year}{2015}).

\bibitem{gasteiger_fast_2020}
\bibinfo{author}{Gasteiger, J.}, \bibinfo{author}{Giri, S.},
  \bibinfo{author}{Margraf, J.~T.} \& \bibinfo{author}{Günnemann, S.}
\newblock \bibinfo{title}{Fast and {Uncertainty}-{Aware} {Directional}
  {Message} {Passing} for {Non}-{Equilibrium} {Molecules}}
  (\bibinfo{year}{2020}).
\newblock \bibinfo{note}{ArXiv:2011.14115 [cs] version: 3}.

\bibitem{reuter_composition_2001}
\bibinfo{author}{Reuter, K.} \& \bibinfo{author}{Scheffler, M.}
\newblock \bibinfo{title}{Composition, structure, and stability of {RuO2}(110)
  as a function of oxygen pressure}.
\newblock \emph{\bibinfo{journal}{Physical Review B}}
  \textbf{\bibinfo{volume}{65}}~(3), \bibinfo{pages}{035406}
  (\bibinfo{year}{2001}).
\newblock \doi{10.1103/PhysRevB.65.035406}.

\bibitem{heifets_density_2007}
\bibinfo{author}{Heifets, E.}, \bibinfo{author}{Ho, J.} \&
  \bibinfo{author}{Merinov, B.}
\newblock \bibinfo{title}{Density functional simulation of the {BaZrO3}(011)
  surface structure}.
\newblock \emph{\bibinfo{journal}{Physical Review B}}
  \textbf{\bibinfo{volume}{75}}~(15), \bibinfo{pages}{155431}
  (\bibinfo{year}{2007}).
\newblock \doi{10.1103/PhysRevB.75.155431}.

\bibitem{kresse_efficient_1996}
\bibinfo{author}{Kresse, G.} \& \bibinfo{author}{Furthmüller, J.}
\newblock \bibinfo{title}{Efficient iterative schemes for ab initio
  total-energy calculations using a plane-wave basis set}.
\newblock \emph{\bibinfo{journal}{Physical Review B}}
  \textbf{\bibinfo{volume}{54}}~(16), \bibinfo{pages}{11169--11186}
  (\bibinfo{year}{1996}).
\newblock \doi{10.1103/PhysRevB.54.11169}.

\bibitem{kresse_ultrasoft_1999}
\bibinfo{author}{Kresse, G.} \& \bibinfo{author}{Joubert, D.}
\newblock \bibinfo{title}{From ultrasoft pseudopotentials to the projector
  augmented-wave method}.
\newblock \emph{\bibinfo{journal}{Physical Review B}}
  \textbf{\bibinfo{volume}{59}}~(3), \bibinfo{pages}{1758--1775}
  (\bibinfo{year}{1999}).
\newblock \doi{10.1103/PhysRevB.59.1758}.

\bibitem{perdew_generalized_1996}
\bibinfo{author}{Perdew, J.~P.}, \bibinfo{author}{Burke, K.} \&
  \bibinfo{author}{Ernzerhof, M.}
\newblock \bibinfo{title}{Generalized {Gradient} {Approximation} {Made}
  {Simple}}.
\newblock \emph{\bibinfo{journal}{Physical Review Letters}}
  \textbf{\bibinfo{volume}{77}}~(18), \bibinfo{pages}{3865--3868}
  (\bibinfo{year}{1996}).
\newblock \doi{10.1103/PhysRevLett.77.3865}.

\bibitem{tadmor_potential_2011}
\bibinfo{author}{Tadmor, E.~B.}, \bibinfo{author}{Elliott, R.~S.},
  \bibinfo{author}{Sethna, J.~P.}, \bibinfo{author}{Miller, R.~E.} \&
  \bibinfo{author}{Becker, C.~A.}
\newblock \bibinfo{title}{The potential of atomistic simulations and the
  knowledgebase of interatomic models}.
\newblock \emph{\bibinfo{journal}{JOM}} \textbf{\bibinfo{volume}{63}}~(7),
  \bibinfo{pages}{17--17} (\bibinfo{year}{2011}).
\newblock \doi{10.1007/s11837-011-0102-6}.

\bibitem{du_data_2023}
\bibinfo{author}{Du, X.}
\newblock \bibinfo{title}{Data for: {Machine} learning-accelerated simulations
  enable automatic surface reconstruction} (\bibinfo{year}{2023}).
\newblock \doi{10.5281/zenodo.7758174}.

\bibitem{du_2023_code}
\bibinfo{author}{Du, X.}
\newblock \bibinfo{title}{learningmatter-mit/surface-sampling}
  (\bibinfo{year}{2023}).
\newblock \doi{10.5281/zenodo.10086398}.

\end{thebibliography}

\clearpage

\begin{thebibliography}{9}
\expandafter\ifx\csname url\endcsname\relax
  \def\url#1{\burl{#1}}\fi
\expandafter\ifx\csname urlprefix\endcsname\relax\def\urlprefix{URL }\fi
\providecommand{\bibinfo}[2]{#2}
\providecommand{\eprint}[2][]{\url{#2}}
\providecommand{\doi}[1]{\url{https://doi.org/#1}}

\bibitem{reuter_composition_2001}
\bibinfo{author}{Reuter, K.} \& \bibinfo{author}{Scheffler, M.}
\newblock \bibinfo{title}{Composition, structure, and stability of \ce{RuO2}(110) as a function of oxygen
  pressure}.
\newblock \emph{\bibinfo{journal}{Physical Review B}}
  \textbf{\bibinfo{volume}{65}}~(3), \bibinfo{pages}{035406}
  (\bibinfo{year}{2001}).
\newblock \doi{10.1103/PhysRevB.65.035406}.

\bibitem{heifets_density_2007}
\bibinfo{author}{Heifets, E.}, \bibinfo{author}{Ho, J.} \&
  \bibinfo{author}{Merinov, B.}
\newblock \bibinfo{title}{Density functional simulation of the \ce{BaZrO3}(011) surface structure}.
\newblock \emph{\bibinfo{journal}{Physical Review B}}
  \textbf{\bibinfo{volume}{75}}~(15), \bibinfo{pages}{155431}
  (\bibinfo{year}{2007}).
\newblock \doi{10.1103/PhysRevB.75.155431}.

\bibitem{heifets_electronic_2007}
\bibinfo{author}{Heifets, E.}, \bibinfo{author}{Piskunov, S.},
  \bibinfo{author}{Kotomin, E.~A.}, \bibinfo{author}{Zhukovskii, Y.~F.} \&
  \bibinfo{author}{Ellis, D.~E.}
\newblock \bibinfo{title}{Electronic structure and thermodynamic stability of
  double-layered \ce{SrTiO3}(001) surfaces: \textit{{Ab} initio}
  simulations}.
\newblock \emph{\bibinfo{journal}{Physical Review B}}
  \textbf{\bibinfo{volume}{75}}~(11), \bibinfo{pages}{115417}
  (\bibinfo{year}{2007}).
\newblock \doi{10.1103/PhysRevB.75.115417}.

\bibitem{thomas_c_allison_nist-janaf_2013}
\bibinfo{author}{Allison, T.~C.}
\newblock \bibinfo{title}{{NIST}-{JANAF} {Thermochemical} {Tables} - {SRD} 13}
  (\bibinfo{year}{2013}).
\newblock \urlprefix\url{https://janaf.nist.gov/}.

\end{thebibliography}
\end{document}